# Nanoparticles with Cubic Symmetry: Classification of Polyhedral Shapes

## Klaus E. Hermann

Theory Dept., Fritz-Haber-Institut der Max-Planck-Gesellschaft,
Faradayweg 4-6, 14195 Berlin, Germany.
Corresponding Author: hermann@fhi-berlin.mpg.de

## Abstract

The shape of crystalline nanoparticles (NP) can often be described by polyhedra with flat facet surfaces. Thus, structural studies of polyhedral bodies can help to describe geometric details of NPs. Here we consider compact polyhedra of cubic point symmetry $O_h$ as simple models. Their surfaces are described by facets with normal vectors along selected directions ($a$, $b$, $c$) together with their symmetry equivalents forming a direction family $\{abc\}$. For given $\{abc\}$ this yields generic polyhedra with up to 48 facets where we focus on polyhedra with facets of $\{abc\} = \{100\}$, $\{110\}$, and $\{111\}$, suggested for metal NPs with cubic lattices. The resulting generic polyhedra, cubic, rhombohedral, and octahedral, can serve for the description of non-generic polyhedra as intersections of corresponding generic species. Their structural properties are shown to be fully determined by only three structure parameters, facet distances $R_{100}$, $R_{110}$, and $R_{111}$ of three types of facets. This provides a phase diagram to completely classify the corresponding $O_h$ symmetry polyhedra. Structural properties of all polyhedra, such as shape, size, and facet geometries, are discussed in analytical and numerical detail with visualization of characteristic examples. The results may be used for respective nanoparticle simulations but also as a repository assisting the interpretation of structures of real compact nanoparticles observed by experiment.

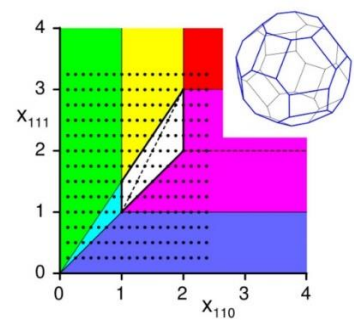

# 1. Introduction

Recently, the detailed characterization of polyhedral bodies, while a subject of mathematical resarch since ancient times [1], has attracted new interest in connection with crystalline nanoparticles (NPs) [2-5]. These particles come in many sizes and polyhedral shapes. Their properties have been explored both experimentally and in theoretical studies due to their exciting physical and chemical behavior, which deviates often from that of corresponding bulk material [2-4]. Examples are applications in medicine [6] or in catalytic chemistry where metal nanoparticles have become ubiquitous [7, 8].

Many metal NPs have been observed in experiments to exhibit polyhedral shape with flat local surface areas (facets) of high atom density, reminiscent of low Miller index planes in corresponding cubic bulk crystals [9, 10]. This can be associated with the geometry of the corresponding crystal lattice suggesting crystalline bulk structure inside the NP. At an atomic scale, the facets at the NP surface join to form edges and corners whose detailed structure can, however, be rather complex. This is due to the discrete distribution of atom positions giving rise to corner capping with microfacets and edge flattening leading to microstrips as discussed earlier [5]. The perturbative effect is even enhanced by local relaxation at the particle surface and as a result of chemical surface reactions.

The overall shape of experimental metal NPs often reminds of compact sections confined by simple polyhedra of $O_h$ symmetry, such as cubes, octahedra, rhombohedra, and others, which may be attributed to the cubic lattice structure of the corresponding bulk metal lattice [5, 9, 10]. Here the analysis of ideal polyhedra with cubic $O_h$ symmetry as NP envelops can be helpful to obtain further insight into possible NP shapes and their classification. Corresponding analytical results of the polyhedral structure allow estimates of NP sizes depending on the number of atoms included together with atom densites in bulk metal. They also give insight into the geometry of possible facets at NP surfaces.

In the present work, extending previous theoretical analyses [5, 11], we focus on geometrical details of polyhedra of cubic $O_h$ symmetry with their symmetry center at the origin of a Cartesian coordinate system. The polyhedral surfaces can be described by facets representing planar sections with normal vectors along selected directions $(a, b, c)$ together with their $O_h$ symmetry equivalents. For metal NPs with cubic lattice geometry, possible facets are observed to represent sections of high density monolayers of the cubic bulk characterized by Miller index families $\{hkl\} = \{100\}, \{110\}$, and $\{111\}$ [10] which seem to be energetically preferred. This suggests polyhedra with facet normal vectors $(a, b, c) = (1, 0, 0)$, $(1, 1, 0)$, and $(1, 1, 1)$ in Cartesian coordinates together with their $O_h$ symmetry equivalents. The analysis reveals different types of generic polyhedra which can serve for the definition of general polyhedra described as intersections of corresponding generic species. Their structural properties, such as shape, size, and surface facets, are shown to be fully determined by only three structure parameters, the facet distances $R_{100}$, $R_{110}$, $R_{111}$. In fact, all polyhedral shapes, independent of size, can already be characterized by only two relative facet distances, such as $x_{110} = R_{110}/R_{100}$ and $x_{111} = R_{111}/R_{100}$ which provides a complete phase diagram of all polyhedral shapes.

We also consider generic polyhedra of $O_h$ symmetry which are confined by facets of one general direction family $\{abc\}$, yielding up to 48 different facet directions. These polyhedra can be used to model metal NPs with higher Miller index facets reflecting sections of stepped and kinked facet surfaces [10]. Clearly, their structural properties are fully described by a facet distance $R_{abc}$ and all components, $a$, $b$, $c$, determining the corresponding facet normal vector family.

All structural results of the present polyhedra are discussed in analytical and numerical detail with visualization [12] of characteristic examples. The different sections are structured identically and presented in parts with very similar phrasing to enable easy comparison. These results allow a full classification of all corresponding polyhedra which may be used as a repository to assist the interpretation of structures of real compact NPs observed by experiment. They can also be useful for corresponding nanoparticle simulations. Also mesoscopic crystallites with internal cubic lattice assuming polyhedral shape may be classified by the present scheme.

Sec. 2 introduces notations and definitions used to characterize polyhedral shape while Sec. 3 discusses many examples of polyhedra, generic and non-generic, in detail. Finally, Sec. 4 summarizes conclusions from the present work. The supplement provides further details to complement results discussed in Sec. 3.

## 2. Notation and Formal Definitions

We consider compact polyhedra of central $O_h$ symmetry confined by finite sections of planes (facets) which can be described by facet normal vectors $\underline{e}_{abc}$ and facet distances $R_{abc}$ from the polyhedral center, resulting in facet vectors

$$\underline{R}_{abc} = R_{abc}\, \underline{e}_{abc}\,, \qquad \underline{e}_{abc} = 1/w\,(a, b, c)\,, \qquad w = \sqrt{(a^2 + b^2 + c^2)}\,, \qquad (1)$$

with $\underline{e}_{abc}$ described by facet indices $a$, $b$, $c$ in Cartesian coordinates relative to the polyhedral center. Due to the polyhedral $O_h$ symmetry each facet normal vector $\underline{e}_{abc}$ implies a number of symmetry equivalents $\underline{e}_{a'b'c'}$ originating from all $O_h$ symmetry operations applied to $\underline{e}_{abc}$. Together with $\underline{e}_{abc}$, this forms a family of symmetry equivalent facet normal vectors, denoted $\underline{e}_{\{abc\}}$ in the following and corresponding to a direction family defined as $\{abc\}$. (Note that in the following we use a short hand notation taken from crystallography [10]: curly brackets $\{\ldots\}$ to indicate facet normal direction families with all members and normal brackets $(\ldots)$ referring to specific directions.)

In the most general case, applying all 48 $O_h$ symmetry operations to a vector $\underline{e}_{abc}$, where $a$, $b$, $c$ are all finite and different from each other yields a direction family $\{abc\}$ of 48 members described by

$$\underline{e}_{\{abc\}} = \;\; 1/w\,(\pm a, \pm b, \pm c),\; 1/w\,(\pm a, \pm c, \pm b),\; 1/w\,(\pm b, \pm a, \pm c),$$
$$1/w\,(\pm b, \pm c, \pm a),\; 1/w\,(\pm c, \pm a, \pm b),\; 1/w\,(\pm c, \pm b, \pm a) \qquad (2a)$$

Special cases resulting in direction families $\{100\}, \{110\}, \{111\}$ are

$$\underline{e}_{\{100\}} = \;\; (\pm 1, 0, 0),\; (0, \pm 1, 0),\; (0, 0, \pm 1), \qquad (2b)$$

$$\underline{e}_{\{110\}} = \;\; 1/\sqrt{2}\,(\pm 1, \pm 1, 0),\; 1/\sqrt{2}\,(\pm 1, 0, \pm 1),\; 1/\sqrt{2}\,(0, \pm 1, \pm 1) \qquad (2c)$$

$$\underline{e}_{\{111\}} = \;\; 1/\sqrt{3}\,(\pm 1, \pm 1, \pm 1) \qquad (2d)$$

which yield smaller families of 6, 12, and 8 members, respectively.

As a result of the $O_h$ symmetry, polyhedral facets appear always as parallel pairs with facet vectors $\pm \underline{R}_{abc}$ on opposite sides of the polyhedron. This leads to polyhedral diameters $D_{abc} = 2\,R_{abc}$ characterizing the size of the polyhedron. Altogether, the most general polyhedra of $O_h$ symmetry can be denoted by

$$P(\underline{R}_{\{abc\}};\; \underline{R}_{\{a'b'c'\}};\; \underline{R}_{\{a''b''c''\}};\; \ldots) =$$
$$P(R_{abc}, \{abc\};\; R_{a'b'c'}, \{a'b'c'\};\; R_{a''b''c''}, \{a''b''c''\};\; \ldots) \qquad (3)$$

depending on the number of different facet types. Here we distinguish between *generic* and *non-generic* species where generic polyhedra are defined by facets of only one direction family $\{abc\}$ whereas non-generic polyhedra include several different direction families as noted in definition (3).

The above notations and definitions will be used in the following discussion. Note that some of the expressions of corner coordinates in Secs. 3.2 use auxiliary parameters $g$, $h$ which are defined separately for each Section.

# 3. Discussion of Example Polyhedra

## 3.1. Generic Polyhedra

As discussed above, generic polyhedra are confined by facets of only one direction family $\{abc\}$ and are denoted P($\underline{R}_{\{abc\}}$). Here the simplest examples are those for $\{abc\}$ = {100}, {110}, and {111} which will be discussed before the general case, which includes also the simple examples, is treated in detail.

### 3.1.1 Cubic Polyhedra P($\underline{R}_{\{100\}}$)

According to (2b), (3), these polyhedra are confined by all 6 {100} facets with facet distances $R_{100}$ which describes, as expected, a *cubic* polyhedron, see Fig. 1a.

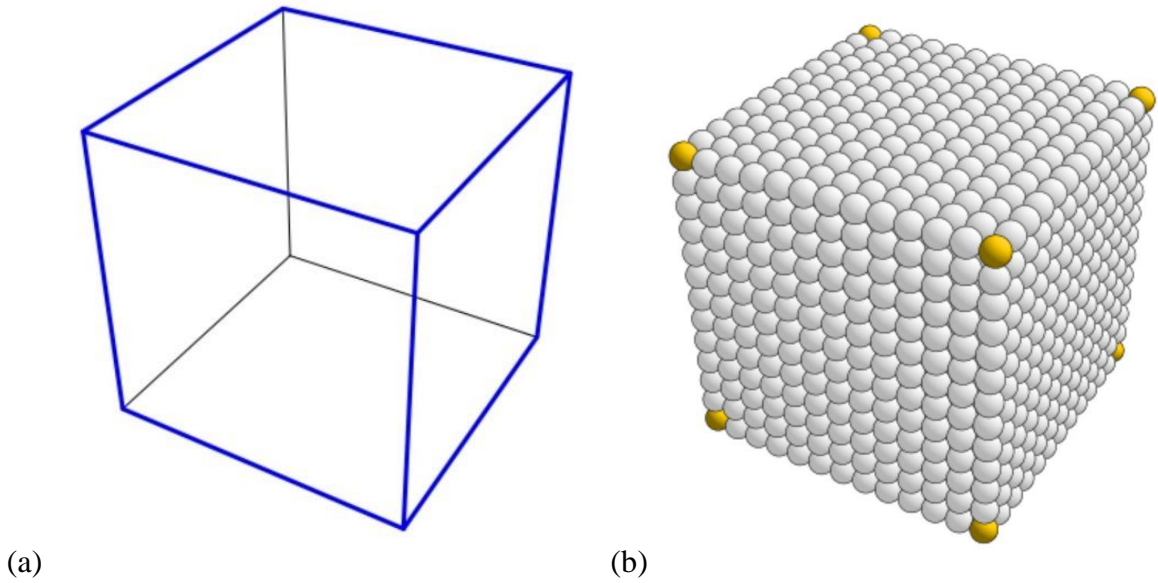

(a)                                     (b)

**Figure 1.** (a) Sketch of generic cubic polyhedron P($\underline{R}_{\{100\}}$) with front facets in blue and back facets in black. (b) Nanoparticle of gray atom balls of a sc crystal section filling the polyhedron, see text. Yellow balls correspond to corners $\underline{C}_{\{111\}}$ which coincide with atom sites.

The 8 polyhedral corners are described by vectors $\underline{C}_{\{111\}}$ relative to the center where in Cartesian coordinates

$$\underline{C}_{\{111\}} = R_{100} (\pm 1, \pm 1, \pm 1) \qquad (4)$$

Euler's polyhedron rule [1] states that

$$N_c + N_f - N_e = 2 \qquad (5)$$

where $N_c$, $N_f$, $N_e$ are the numbers of corners, facets, and edges, respectively, of a convex polyhedron. This yields P($\underline{R}_{\{100\}}$) with corners connected by 12 (8 + 6 - 2) edges, see Fig.1a.

An analysis shows that all 6 facets are of the same square shape where each {100} facet extends between four adjacent corners $\underline{C}_{\{111\}}$, such as $\underline{C}_{(111)}$, $\underline{C}_{(-111)}$, $\underline{C}_{(-1-11)}$, $\underline{C}_{(1-11)}$. The resulting four edges connect corners, such as $\underline{C}_{(111)}$ with $\underline{C}_{(-111)}$, at distances $d_{a1}$ given by

$$d_{a1} = 2R_{100} \tag{6}$$

The largest distance from the polyhedral center to its surface along $(abc)$ directions, $s_{abc}(R_{100})$, is given by

$$s_{100}(R_{100}) = R_{100} \tag{7a}$$

$$s_{110}(R_{100}) = \sqrt{2}\, R_{100} \tag{7b}$$

$$s_{111}(R_{100}) = \sqrt{3}\, R_{100} \tag{7c}$$

Further, the area of each facet is given by $F_0$ with

$$F_0 = |\, (\underline{C}_{(-111)} - \underline{C}_{(111)}) \times (\underline{C}_{(1-11)} - \underline{C}_{(111)})\, | = 4\, R_{100}^2 \tag{8}$$

Thus, the total facet surface, $F_{surf}$ (sum over all facet areas) and the volume $V_{tot}$ of the polyhedron are given by

$$F_{surf} = 6\, F_0 = 24\, R_{100}^2 \tag{9}$$

$$V_{tot} = F_{surf}\, R_{100}\, / \, 3 = 8\, R_{100}^3 \tag{10}$$

Fig. 1b shows a nanoparticle of atom balls representing a simple cubic (sc) crystal section where a polyhedron $P(\underline{R}_{\{100\}})$ serves as envelope with its corners $\underline{C}_{\{111\}}$ coinciding with atom sites.

### 3.1.2 Rhombohedral Polyhedra $P(\underline{R}_{\{110\}})$

According to (2c), (3), these polyhedra are confined by all 12 {110} facets with facet distances $R_{110}$ which describes a *rhombohedral* polyhedron, see Fig. 2a.

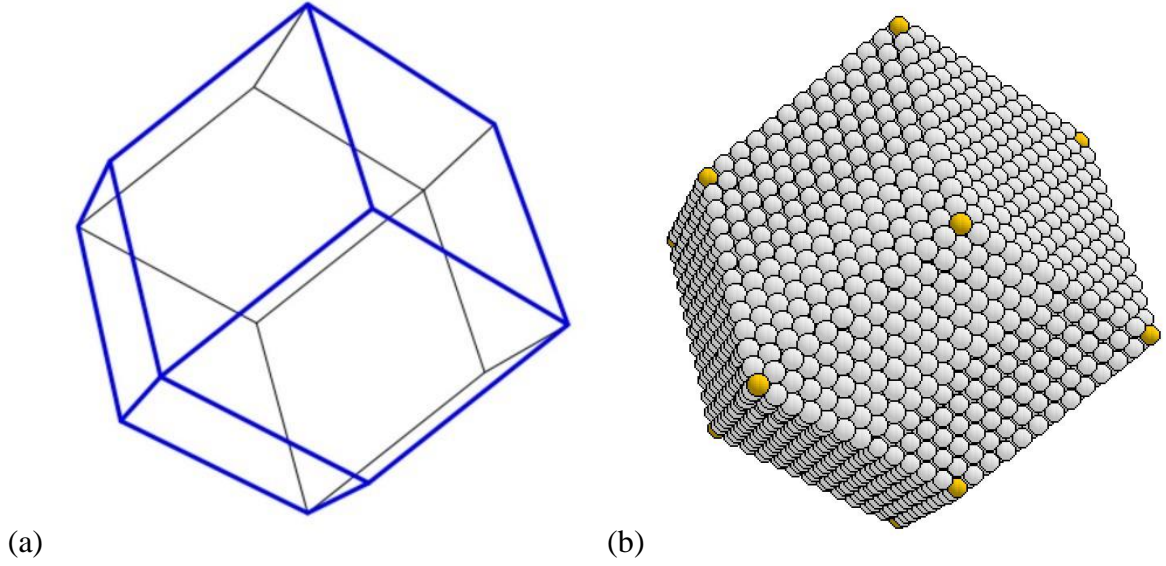

(a)                                             (b)

**Figure 2.** (a) Sketch of generic rhombohedral polyhedron P($\underline{R}_{\{110\}}$) with front facets in blue and back facets in black. (b) Nanoparticle of gray atom balls of a bcc lattice section filling the polyhedron. Yellow balls correspond to corners $\underline{C}_{\{100\}}$ and $\underline{C}_{\{111\}}$ which coincide with atom sites.

The 14 polyhedral corners fall in two groups of 6 and 8 each, described by vectors $\underline{C}_{\{100\}}$ and $\underline{C}_{\{111\}}$ relative to the center, where in Cartesian coordinates

$$\underline{C}_{\{100\}} = \sqrt{2}\, R_{110}\, (\pm 1, 0, 0)\,, \quad = \sqrt{2}\, R_{110}\, (0, \pm 1, 0)\,, \quad = \sqrt{2}\, R_{110}\, (0, 0, \pm 1) \qquad (11a)$$

$$\underline{C}_{\{111\}} = 1/\sqrt{2}\, R_{110}\, (\pm 1, \pm 1, \pm 1) \qquad (11b)$$

With P($\underline{R}_{\{110\}}$) yielding 12 facets and 14 corners the number of its polyhedral edges amounts to 24 according to (5), see Fig. 2a.

An analysis shows that all 12 facets are of the same rhombic shape where each {110} facet extends between adjacent corners $\underline{C}_{\{100\}}$ and $\underline{C}_{\{111\}}$, such as $\underline{C}_{(100)}$, $\underline{C}_{(111)}$, $\underline{C}_{(001)}$, $\underline{C}_{(1\text{-}11)}$. The resulting four edges connect corners, such as $\underline{C}_{(100)}$ with $\underline{C}_{(111)}$, at distances $d_{b1}$ given by

$$d_{b1} = \sqrt{(3/2)}\, R_{110} \qquad (12)$$

Thus, the polyhedron can be described as a rhombic dodecahedron reminding of the shape of Wigner-Seitz cells of the face-centered cubic crystal lattice [13].

The largest distance from the polyhedral center to its surface along (*abc*) directions, $s_{abc}(R_{110})$, is given by

$$s_{100}(R_{110}) = \sqrt{2}\, R_{110} \qquad (13a)$$

$$s_{110}(R_{110}) = R_{110} \qquad (13b)$$

$$s_{111}(R_{110}) = \sqrt{(3/2)}\, R_{110} \qquad (13c)$$

Further, the area of each facet is given by $F_0$ with

$$F_0 = |(\underline{C}_{(001)} - \underline{C}_{(111)}) \times (\underline{C}_{(100)} - \underline{C}_{(111)})| = \sqrt{2}\, R_{110}^2 \qquad (14)$$

Thus, the total facet surface, $F_{surf}$ (sum over all facet areas) and the volume $V_{tot}$ of the polyhedron are given by

$$F_{surf} = 12\, F_0 = 12\, \sqrt{2}\, R_{110}^2 \qquad (15)$$

$$V_{tot} = F_{surf}\, R_{110}\, / \, 3 = 4\, \sqrt{2}\, R_{110}^3 \qquad (16)$$

Fig. 2b shows a nanoparticle of atom balls representing a body-centered cubic (bcc) crystal section where a polyhedron P($\underline{R}_{\{110\}}$) serves as envelope with its corners $\underline{C}_{\{100\}}$ and $\underline{C}_{\{111\}}$ coinciding with atom sites.

### 3.1.3 Octahedral Polyhedra P($\underline{R}_{\{111\}}$)

According to (2d), (3), these polyhedra are confined by all 8 {111} facets with facet distances $R_{111}$ which describes an *octahedral* polyhedron, see Fig. 3a.

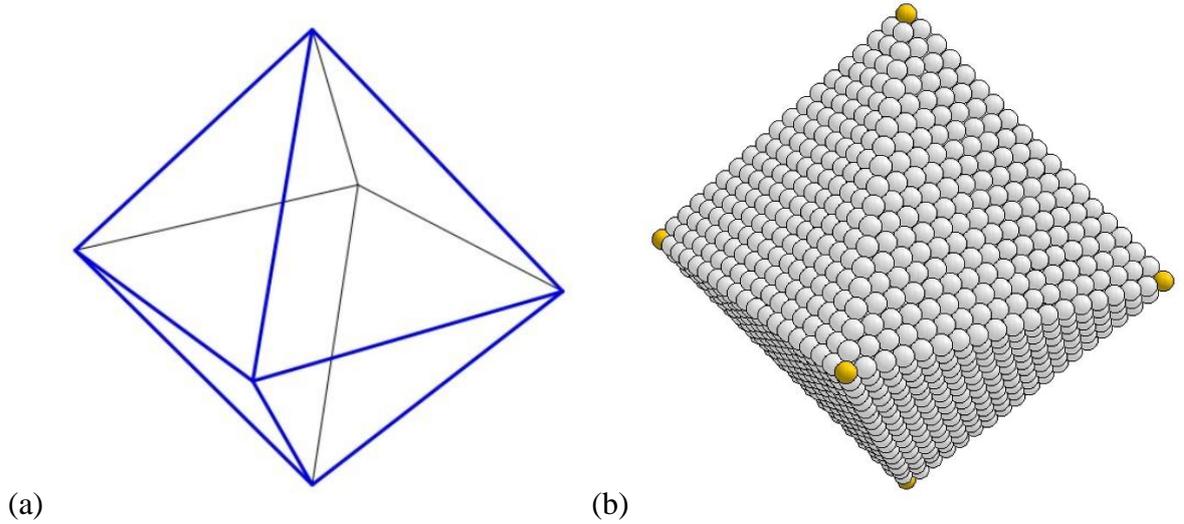

(a)　　　　　　　　　　　　　　　　(b)

**Figure 3.** (a) Sketch of generic octahedral polyhedron P($\underline{R}_{\{111\}}$) with front facets in blue and back facets in black. (b) Nanoparticle of gray atom balls of an fcc lattice section filling the polyhedron. Yellow balls correspond to corners $\underline{C}_{\{100\}}$ which coincide with atom sites.

The 6 polyhedral corners are described by vectors $\underline{C}_{\{100\}}$ relative to the center where in Cartesian coordinates

$$\underline{C}_{\{100\}} = \sqrt{3}\, R_{111}\, (\pm 1, 0, 0)\,, \quad = \sqrt{3}\, R_{111}\, (0, \pm 1, 0)\,, \quad = \sqrt{3}\, R_{111}\, (0, 0, \pm 1) \qquad (17)$$

With P($\underline{R}_{\{110\}}$) yielding 8 facets and 6 corners the number of its polyhedral edges amounts to 12 according to (5), see Fig. 3a.

An analysis shows that all 8 facets are of the same equilateral triangular shape where each $\{111\}$ facet extends between adjacent corners $\underline{C}_{\{100\}}$, such as $\underline{C}_{(100)}$, $\underline{C}_{(010)}$, $\underline{C}_{(001)}$. The resulting three edges connect corners, such as $\underline{C}_{(010)}$ with $\underline{C}_{(100)}$, at distances $d_{c1}$ given by

$$d_{c1} = \sqrt{6}\, R_{111} \qquad (18)$$

The largest distance from the polyhedral center to its surface along $(abc)$ directions, $s_{abc}(R_{111})$, is given by

$$s_{100}(R_{111}) = \sqrt{3}\, R_{111} \qquad (19a)$$

$$s_{110}(R_{111}) = \sqrt{(3/2)}\, R_{111} \qquad (19b)$$

$$s_{111}(R_{111}) = R_{111} \qquad (19c)$$

Further, the area of each facet is given by $F_0$ with

$$F_0 = 1/2 \, | \, (\underline{C}_{(100)} - \underline{C}_{(001)}) \times (\underline{C}_{(010)} - \underline{C}_{(001)}) \, | \, = \, (3/2)\sqrt{3}\, R_{111}{}^2 \qquad (20)$$

Thus, the total facet surface, $F_{surf}$ (sum over all facet areas) and the volume $V_{tot}$ of the polyhedron are given by

$$F_{surf} = 8\, F_0 = 12\sqrt{3}\, R_{111}{}^2 \qquad (21)$$

$$V_{tot} = F_{surf}\, R_{111} \, / \, 3 = 4\sqrt{3}\, R_{111}{}^3 \qquad (22)$$

Fig. 3b shows a nanoparticle of atom balls representing a face-centerd cubic (fcc) crystal section where a polyhedron P($\underline{R}_{\{111\}}$) serves as envelope with its corners $\underline{C}_{\{100\}}$ coinciding with atom sites.

### 3.1.4 Polyhedra P($\underline{R}_{\{abc\}}$)

According to (2a), (3), these polyhedra are confined by up to 48 $\{abc\}$ facets with facet distances $R_{abc}$. If values of $a$, $b$, $c$ coincide or equal zero, different facets can join to yield larger faces and the number of facet normal vectors decreases to 24, 12, 8, or 6 facets as discussed in Sec .S.1 of the Supplement. Here we focus on the general case of $a > b > c > 0$ which results in 48 different $\{abc\}$ facets, see Fig. 4a.

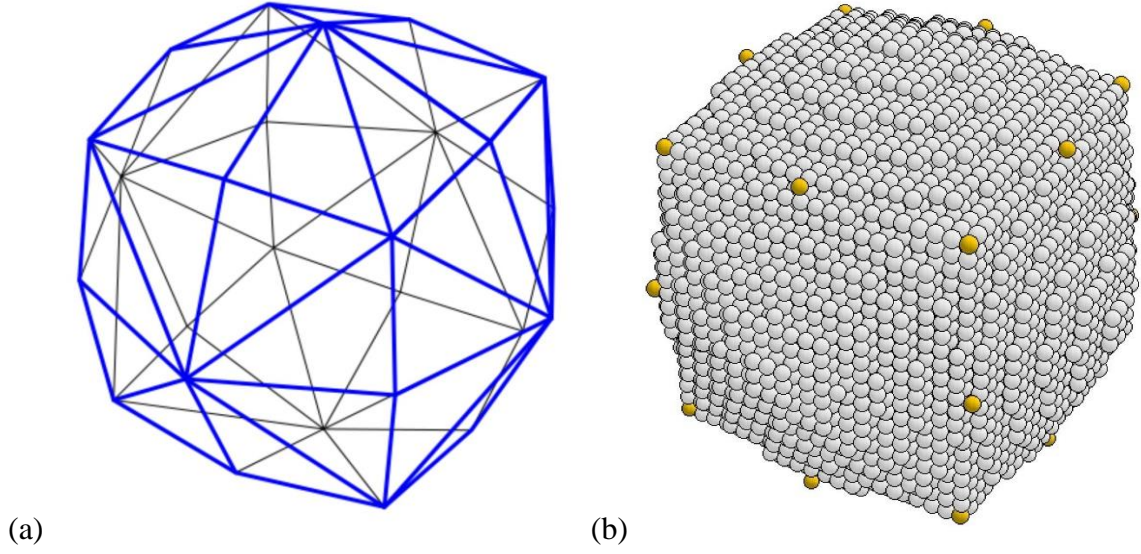

(a)                                      (b)

**Figure 4.** (a) Sketch of polyhedron P($\underline{R}_{\{abc\}}$) for $a = 7$, $b = 3$, c = 1 with front facets in blue and back facets in black. (b) Nanoparticle of gray atom balls of an sc lattice section filling the polyhedron. Yellow balls correspond to corners $\underline{C}_{\{110\}}$ and $\underline{C}_{\{111\}}$ which coincide with atom sites.

As a result of the overall $O_h$ symmetry of the P($\underline{R}_{\{abc\}}$) polyhedron, its corners can appear only along selected directions from the center given by vectors $\underline{e}_{\{100\}}$, $\underline{e}_{\{110\}}$, $\underline{e}_{\{111\}}$ according to (2b), (2c), (2d). This yields possible corner vectors

$$\underline{C}_{\{hkl\}} = p_{hkl} \, \underline{e}_{\{hkl\}} \, , \qquad \{hkl\} = \{100\}, \{110\}, \{111\} \tag{23}$$

where $\underline{C}_{\{hkl\}}$ must point to several joining $\{abc\}$ facets. This requires that

$$\underline{C}_{\{hkl\}} \, \underline{e}_{\{abc\}} = p \, (\underline{e}_{\{hkl\}} \, \underline{e}_{\{abc\}}) = R_{abc} \qquad p_{hkl} = R_{abc} \, / \, (\underline{e}_{\{hkl\}} \, \underline{e}_{\{abc\}}) \tag{24}$$

Together with (2a), (2b), (2c), (2d) we obtain

$$\underline{e}_{\{100\}} \, \underline{e}_{\{abc\}} = a/w \qquad\qquad \underline{C}_{\{100\}} = R_{abc} \, w/a \, \underline{e}_{\{100\}} \tag{25a}$$

$$\underline{e}_{\{110\}} \, \underline{e}_{\{abc\}} = (a+b) \, /(\sqrt{2} \, w) \qquad \underline{C}_{\{110\}} = R_{abc} \, w \, \sqrt{2}/(a+b) \, \underline{e}_{\{110\}} \tag{25b}$$

$$\underline{e}_{\{111\}} \, \underline{e}_{\{abc\}} = (a+b+c) \, /(\sqrt{3} \, w) \qquad \underline{C}_{\{111\}} = R_{abc} \, w \, \sqrt{3}/(a+b+c) \, \underline{e}_{\{111\}} \tag{25c}$$

describing, altogether, 26 different corners. With P($\underline{R}_{\{abc\}}$) yielding 48 facets and 26 corners the number of its polyhedral edges amounts to 72 according to (5), see Fig.4a.

An analysis shows that all all 48 $\{abc\}$ facets are of triangular shape. There are two sets of 24 identical facets each where the facets of the second set are obtained as mirror images of those of the first. Each facet triangle extends between adjacent corners $\underline{C}_{\{100\}}$, $\underline{C}_{\{110\}}$, and $\underline{C}_{\{111\}}$, such as $\underline{C}_{\langle 100\rangle}$, $\underline{C}_{\langle 110\rangle}$, $\underline{C}_{\langle 111\rangle}$. The resulting three edges connect corners $\underline{C}_{\{100\}}$ with $\underline{C}_{\{110\}}$, $\underline{C}_{\{100\}}$ with $\underline{C}_{\{111\}}$, and $\underline{C}_{\{110\}}$ with $\underline{C}_{\{111\}}$, at different distances $d_{d1}$, $d_{d2}$, and $d_{d3}$ given by

$$d_{d1} = R_{abc} \, w \, \sqrt{(a^2 + b^2)} \, / \, [a \, (a + b)] \tag{26a}$$

$$d_{d2} = R_{abc} \, w \, \sqrt{[2a^2 + (b + c)^2)]} \, / \, [a \, (a + b + c)] \tag{26b}$$

$$d_{d3} = R_{abc} \, w \, \sqrt{[(a + b)^2 + 2c^2)]} \, / \, [(a + b) \, (a + b + c)] \tag{26c}$$

Further, the area of each facet is given by $F_0$ with

$$
\begin{aligned}
F_0 \; &= \; 1/2 \, | \, (\underline{C}_{(110)} - \underline{C}_{(100)}) \times (\underline{C}_{(111)} - \underline{C}_{(100)}) \, | \\
&= \; 1/2 \, (R_{abc} \, w)^2 \, w \, / \, [a \, (a + b) \, (a + b + c)] \tag{27}
\end{aligned}
$$

Thus, the total facet surface, $F_{surf}$ (sum over all facet areas) and the volume $V_{tot}$ of the polyhedron are given by

$$F_{surf} \; = \; 48 \, F_0 \; = \; 24 \, (R_{abc} \, w)^2 \, w \, / \, [a \, (a + b) \, (a + b + c)] \tag{28}$$

$$V_{tot} \; = \; F_{surf} \, R_{abc} \, / \, 3 \; = \; 8 \, (R_{abc} \, w)^3 \, / \, [a \, (a + b) \, (a + b + c)] \tag{29}$$

Fig. 4b shows a nanoparticle of atom balls representing a simple cubic (sc) crystal section where a polyhedron P($\underline{R}_{\{abc\}}$) serves as envelope with its corners $\underline{C}_{\{110\}}$ and $\underline{C}_{\{111\}}$ coinciding with atom sites. Note that polyhedral corners $\underline{C}_{\{100\}}$ do not appear due to the discrete distribution of the atom sites. The figure also illustrates the stepped/kinked structure of the different facet areas.

Other polyhedra P($\underline{R}_{\{a'b'c'\}}$) of $O_h$ symmetry, where components $a'$, $b'$, $c'$ are not subject to constraints $a' > b' > c' > 0$ imposed in this Section, can be treated completely analogous to the present discussion. First, we note that if $a'$, $b'$, $c'$ are permutations of $a$, $b$, $c$ the polyhedra P($\underline{R}_{\{a'b'c'\}}$) and P($\underline{R}_{\{abc\}}$) are identical in shape. Second, mirror symmetry requires that if any of the components $a'$, $b'$, $c'$ is negative it can be replaced by the corresponding positive value without affecting the polyhedron shape. Thus, $a'$, $b'$, $c'$ can always be regrouped and its component values inverted to yield $a$, $b$, $c$ with $a \geq b \geq c \geq 0$ while conserving the polyhedron shape. So far, we focussed on polyhedra P($\underline{R}_{\{abc\}}$) where equality and zero values of the components $a$, $b$, $c$ are ignored. However, all other cases, including P($\underline{R}_{\{100\}}$), P($\underline{R}_{\{110\}}$), and P($\underline{R}_{\{111\}}$) of Secs. 3.1.1-3, are discussed in Sec. S.1 of the Supplement.

## 3.2. Non-generic Polyhedra

Non-generic polyhedra of $O_h$ symmetry P($\underline{R}_{\{abc\}}$; $\underline{R}_{\{a'b'c'\}}$; …) show facets with orientations of more than one family of facet vectors $\underline{R}_{\{abc\}}$. This can be considered as combining confinements of corresponding different generic polyhedra P($\underline{R}_{\{abc\}}$), discussed in Sec. 3.1, which share their symmetry center. Thus, non-generic polyhedra represent mutual intersections of more than one generic polyhedron, where one cuts corners and edges of the other(s) to form additional facets.

In this section we restrict ourselves to non-generic polyhedra with up to three selected generic polyhedra, cubic P($\underline{R}_{\{100\}}$), rhombohedral P($\underline{R}_{\{110\}}$), and octahedral P($\underline{R}_{\{111\}}$) which offer {100}, {110}, as well as {111} facets with facet distances $R_{100}$, $R_{110}$, and $R_{111}$. This choice is motivated by the structure of ideal cubic metal nanoparticles whose bulk atoms form sections of cubic crystals (simple, face-, and body-centered) and where corresponding facets of {100}, {110}, and {111} families reflect crystal monolayers of highest atom density [10].

The corresponding facet distances $R_{100}$, $R_{110}$, and $R_{111}$ can be considered as structure parameters, defining the present non-generic polyhedra, and their relations with each other determine the polyhedral shape. In the following, we discuss the three types of polyhedra, which combine two generic polyhedra each, i.e. P($\underline{R}_{\{100\}}$; $\underline{R}_{\{110\}}$), P($\underline{R}_{\{100\}}$; $\underline{R}_{\{111\}}$), and P($\underline{R}_{\{110\}}$; $\underline{R}_{\{111\}}$) in Secs. 3.2.1-3, before we consider the most general case of polyhedra as intersections of three generic polyhedra, P($\underline{R}_{\{100\}}$; $\underline{R}_{\{110\}}$; $\underline{R}_{\{111\}}$), in Sec. 3.2.4.

### 3.2.1. Cubo-rhombic Polyhedra P($\underline{R}_{\{100\}}$; $\underline{R}_{\{110\}}$)

Non-generic polyhedra P($\underline{R}_{\{100\}}$; $\underline{R}_{\{110\}}$), denoted *cubo-rhombic*, represent intersections of two generic polyhedra, cubic P($\underline{R}_{\{100\}}$) and rhombohedral P($\underline{R}_{\{110\}}$), see Fig. 5. If the edges of the cubic polyhedron P($\underline{R}_{\{100\}}$) lie inside the rhombohedral polyhedron P($\underline{R}_{\{110\}}$), the resulting combination P($\underline{R}_{\{100\}}$; $\underline{R}_{\{110\}}$) will be generic cubic. This requires

$$s_{110}(R_{100}) \leq s_{110}(R_{110}) \tag{30}$$

and according to (7b), (13b)

$$R_{110} \geq \sqrt{2}\, R_{100} \tag{31}$$

On the other hand, if the corners of the rhombohedral polyhedron P($\underline{R}_{\{110\}}$) lie inside the cubic polyhedron P($\underline{R}_{\{100\}}$), the resulting combination P($\underline{R}_{\{100\}}$; $\underline{R}_{\{110\}}$) will be generic rhombohedral. This requires

$$s_{100}(R_{110}) \leq s_{100}(R_{100}) \tag{32}$$

and according to (7a), (13a)

$$R_{110} \leq 1/\sqrt{2}\, R_{100} \tag{33}$$

Thus, the two generic polyhedra intersect and yield a true polyhedron P($\underline{R}_{\{100\}}$; $\underline{R}_{\{110\}}$) with both {100} and {110} facets only for facet distances $R_{100}$, $R_{110}$ with

$$1/\sqrt{2}\, R_{100} < R_{110} < \sqrt{2}\, R_{100} \tag{34}$$

while P($\underline{R}_{\{100\}}$; $\underline{R}_{\{110\}}$) is generic cubic for $R_{110} \geq \sqrt{2} \, R_{100}$ and generic rhombohedral for $R_{110} \leq 1/\sqrt{2} \, R_{100}$. As a consequence, generic polyhedra P($\underline{R}_{\{100\}}$) and P($\underline{R}_{\{110\}}$) can be described alternatively by non-generic P($\underline{R}_{\{100\}}$; $\underline{R}_{\{110\}}$) where

$$P(\underline{R}_{\{100\}}) \ = \ P(\underline{R}_{\{100\}}; \underline{R}_{\{110\}}) \ \text{with} \ R_{110} \geq \sqrt{2} \, R_{100} \qquad \text{(cubic)} \qquad (35a)$$

$$P(\underline{R}_{\{110\}}) \ = \ P(\underline{R}_{\{100\}}; \underline{R}_{\{110\}}) \ \text{with} \ R_{100} \geq \sqrt{2} \, R_{110} \qquad \text{(rhombohedral)} \qquad (35b)$$

The surfaces of general cubo-rhombic polyhedra P($\underline{R}_{\{100\}}$; $\underline{R}_{\{110\}}$) exhibit 6 {100} facets and 12 {110} facets as shown in Fig. 5.

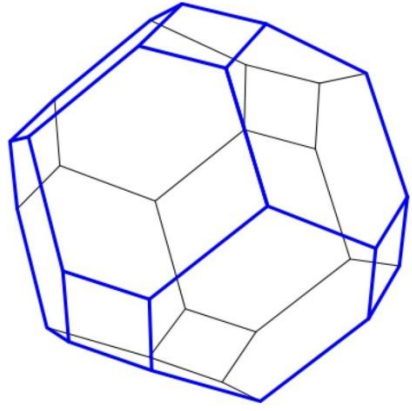

**Figure 5.** Sketch of cubo-rhombic polyhedron P($\underline{R}_{\{100\}}$; $\underline{R}_{\{110\}}$),
$R_{110}/R_{100} = 0.864$, with front facets in blue and back facets in black.

There are 32 polyhedral corners which can be evaluated by methods described in Sec. S.3 of the Supplement. They fall into two groups of 24 and 8 corners each, described by vectors $\underline{C}_{\{1hh\}}$ and $\underline{C}_{\{111\}}$ relative to the center, where in Cartesian coordinates

$$\underline{C}_{\{1hh\}} \ = \ R_{100} \, (\pm 1, \pm h, \pm h) \ , \quad = \ R_{100} \, (\pm h, \pm 1, \pm h) \ , \ = \ R_{100} \, (\pm h, \pm h, \pm 1) \qquad (36)$$

$$\underline{C}_{\{111\}} \ = \ 1/\sqrt{2} \, R_{110} \, (\pm 1, \pm 1, \pm 1) \ , \qquad h \ = \ \sqrt{2} \, R_{110}/R_{100} - 1 \ , \quad 0 \leq h \leq 1$$

The 6 {100} facets are of the same square shape where each facet extends between four adjacent corners $\underline{C}_{\{1hh\}}$, such as $\underline{C}_{(1hh)}$, $\underline{C}_{(1-hh)}$, $\underline{C}_{(1-h-h)}$, $\underline{C}_{(1h-h)}$. The resulting four edges connect corners, such as $\underline{C}_{(1hh)}$ with $\underline{C}_{(1-hh)}$, at distances $d_{e1}$ given by

$$d_{e1} \ = \ 2 \, (\sqrt{2} \, R_{110} - R_{100}) \qquad (37)$$

The 12 {110} facets are of the same hexagonal shape where each facet extends between six adjacent corners $\underline{C}_{\{1hh\}}$ and $\underline{C}_{\{111\}}$, such as $\underline{C}_{(1hh)}$, $\underline{C}_{(1-hh)}$, $\underline{C}_{(1-11)}$, $\underline{C}_{(h-h1)}$, $\underline{C}_{(hh1)}$, $\underline{C}_{(111)}$. Of the resulting six edges two connect corners, such as $\underline{C}_{(1hh)}$ with $\underline{C}_{(1-hh)}$, at distances $d_{e1}$ according to (37) while four connect corners, such as $\underline{C}_{(1hh)}$ with $\underline{C}_{(111)}$, at distances $d_{e2}$ given by

$$d_{e2} \ = \ \sqrt{(3/2)} \, (\sqrt{2} \, R_{100} - R_{110}) \qquad (38)$$

The largest distance from the polyhedral center to its surface along (*abc*) directions, $s_{abc}(R_{100}, R_{110})$, is given by

$$s_{100}(R_{100}, R_{110}) = R_{100} \tag{39a}$$

$$s_{110}(R_{100}, R_{110}) = R_{110} \tag{39b}$$

$$s_{111}(R_{100}, R_{110}) = \sqrt{(3/2)}\, R_{110} \tag{39c}$$

Further, the area of each square {100} facet is given by $F_0$ where with (37)

$$F_0 = 4\,(\sqrt{2}\, R_{110} - R_{100})^2 \tag{40}$$

and of each hexagonal {110} facet by $F_1$ where with (14)

$$F_1 = \sqrt{2}\,(\sqrt{2}\, R_{100} - R_{110})\,(3\, R_{110} - \sqrt{2}\, R_{100}) \tag{41}$$

This yields the total facet surface, $F_{surf}$ (sum over all facet areas) and the volume $V_{tot}$ of the polyhedron according to

$$F_{surf} = 6\, F_0 + 12\, F_1 \tag{42}$$

$$V_{tot} = (\, 6\, F_0\, R_{100} + 12\, F_1\, R_{110}\, ) \, / \, 3 \tag{43}$$

### 3.2.2. Cubo-octahedral Polyhedra P($\underline{R}_{\{100\}}$; $\underline{R}_{\{111\}}$)

Non-generic polyhedra P($\underline{R}_{\{100\}}$; $\underline{R}_{\{111\}}$), denoted *cubo-octahedral*, represent intersections of two generic polyhedra, cubic P($\underline{R}_{\{100\}}$) and octahedral P($\underline{R}_{\{111\}}$), see Fig. 5. If the corners of the cubic polyhedron P($\underline{R}_{\{100\}}$) lie inside the octahedral polyhedron P($\underline{R}_{\{111\}}$), the resulting combination P($\underline{R}_{\{100\}}$; $\underline{R}_{\{111\}}$) will be generic cubic. This requires

$$s_{111}(R_{100}) \leq s_{111}(R_{111}) \tag{44}$$

and according to (7c), (19c)

$$R_{111} \geq \sqrt{3}\, R_{100} \tag{45}$$

On the other hand, if the corners of the octahedral polyhedron P($\underline{R}_{\{111\}}$) lie inside the cubic polyhedron P($\underline{R}_{\{100\}}$), the resulting combination P($\underline{R}_{\{100\}}$; $\underline{R}_{\{111\}}$) will be generic octahedral. This requires

$$s_{100}(R_{111}) \leq s_{100}(R_{100}) \tag{46}$$

and according to (7a), (19a)

$$R_{111} \leq 1/\sqrt{3}\, R_{100} \tag{47}$$

Thus, the two generic polyhedra intersect and yield a true polyhedron P($\underline{R}_{\{100\}}$; $\underline{R}_{\{111\}}$) with both {100} and {111} facets only for facet distances $R_{100}$, $R_{111}$ with

$$1/\sqrt{3}\, R_{100} < R_{111} < \sqrt{3}\, R_{100} \tag{48}$$

while P($\underline{R}_{\{100\}}$; $\underline{R}_{\{111\}}$) is generic cubic for $R_{111} \geq \sqrt{3}\, R_{100}$ and generic octahedral for $R_{111} \leq 1/\sqrt{3}\, R_{100}$. As a consequence, generic polyhedra P($\underline{R}_{\{100\}}$) and P($\underline{R}_{\{111\}}$) can be described alternatively by non-generic P($\underline{R}_{\{100\}}$; $\underline{R}_{\{111\}}$) where

$$P(\underline{R}_{\{100\}}) \;=\; P(\underline{R}_{\{100\}}; \underline{R}_{\{111\}}) \;\text{ with }\; R_{111} \geq \sqrt{3}\, R_{100} \qquad \text{cubic} \tag{49a}$$

$$P(\underline{R}_{\{111\}}) \;=\; P(\underline{R}_{\{100\}}; \underline{R}_{\{111\}}) \;\text{ with }\; R_{100} \geq \sqrt{3}\, R_{111} \qquad \text{octahedral} \tag{49b}$$

The surfaces of general cubo-rhombic polyhedra P($\underline{R}_{\{100\}}$; $\underline{R}_{\{111\}}$) exhibit 6 {100} facets and 8 {111} facets as shown in Fig. 6. Amongst the intersecting species according to (48) we can distinguish between *truncated octahedral* and *truncated cubic* with *cuboctahedral* polyhedra separating where

$$1/\sqrt{3}\, R_{100} < R_{111} < 2/\sqrt{3}\, R_{100} \qquad \text{truncated octahedral} \tag{50a}$$

$$2/\sqrt{3}\, R_{100} < R_{111} < \sqrt{3}\, R_{100} \qquad \text{truncated cubic} \tag{50b}$$

$$R_{111} \;=\; 2/\sqrt{3}\, R_{100} \qquad \text{cuboctahedral} \tag{50c}$$

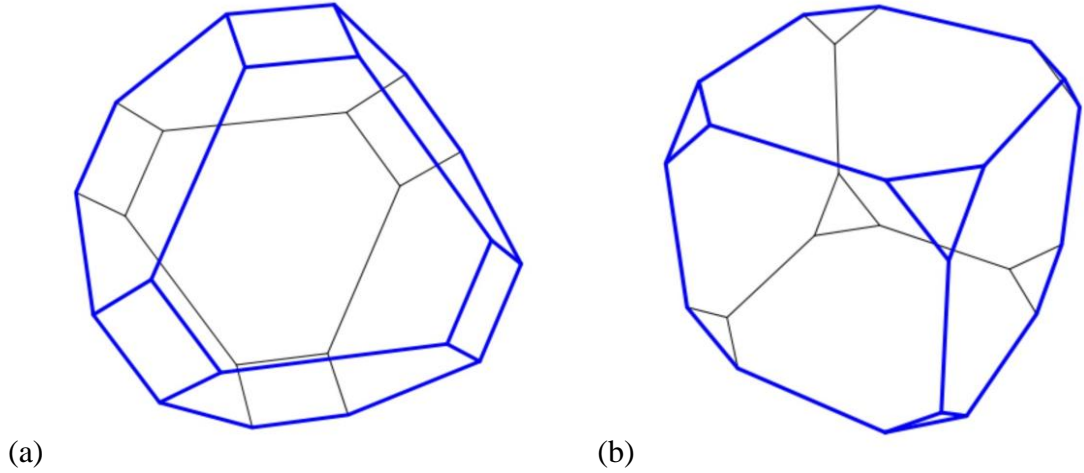

(a)                         (b)

**Figure 6.** Sketch of cubo-octahedral polyhedra P($\underline{R}_{\{100\}}$; $\underline{R}_{\{111\}}$) with front facets in blue and back facets in black, (a) truncated octahedral, $R_{111}/R_{100} = 0.770$, (b) truncated cubic type, $R_{111}/R_{100} = 1.501$.

The surfaces of **truncated octahedral** polyhedra P($\underline{R}_{\{100\}}$; $\underline{R}_{\{111\}}$) with $1 < \sqrt{3} R_{111}/R_{100} < 2$ exhibit 6 {100} facets and 8 {111} facets as shown in Fig. 6a and there are 24 polyhedral corners which can be evaluated by methods described in Sec. S.3 of the Supplement. They are described by vectors $\underline{C}_{\{1h0\}}$ relative to the center, where in Cartesian coordinates

$$\underline{C}_{\{1h0\}} \;=\; R_{100}\,(\pm 1,\, \pm h,\, 0)\, ,\; =\; R_{100}\,(\pm h,\, 0,\, \pm 1)\, ,\; =\; R_{100}\,(0,\, \pm 1,\, \pm h) \tag{51}$$

$$=\; R_{100}\,(\pm 1,\, 0,\, \pm h)\, ,\; =\; R_{100}\,(0,\, \pm h,\, \pm 1)\, ,\; =\; R_{100}\,(\pm h,\, \pm 1,\, 0)$$

$$h \;=\; \sqrt{3}\, R_{111}/R_{100} - 1\, , \quad 0 \leq h \leq 1$$

The 6 {100} facets are of the same square shape where each facet extends between four adjacent corners $\underline{C}_{\{1h0\}}$, such as $\underline{C}_{(1h0)}$, $\underline{C}_{(10h)}$, $\underline{C}_{(1-h0)}$, $\underline{C}_{(10-h)}$. The resulting four edges connect corners, such as $\underline{C}_{(1h0)}$ with $\underline{C}_{(10h)}$, at distances $d_{\mathrm{f1}}$ given by

$$d_{\mathrm{f1}} = \sqrt{2} \, (\sqrt{3} R_{111} - R_{100}) \qquad (52)$$

The 8 {111} facets are of the same hexagonal shape where each facet extends between adjacent corners $\underline{C}_{\{1h0\}}$ and $\underline{C}_{\{111\}}$, such as $\underline{C}_{(1h0)}$, $\underline{C}_{(h10)}$, $\underline{C}_{(01h)}$, $\underline{C}_{(0h1)}$, $\underline{C}_{(h01)}$, $\underline{C}_{(10h)}$. Of the resulting six alternating edges three connect corners, such as $\underline{C}_{(1h0)}$ with $\underline{C}_{(10h)}$, at distances $d_{\mathrm{f1}}$ according to (52) while three connect corners, such as $\underline{C}_{(1h0)}$ with $\underline{C}_{(h10)}$, at distances $d_{\mathrm{f2}}$ given by

$$d_{\mathrm{f2}} = \sqrt{2} \, (2 R_{100} - \sqrt{3} R_{111}) \qquad (53)$$

For $h = 1/2$ all 6 edge lengths are equal leading to regular hexagonal {111} facets. As a result, the polyhedron is reminiscent of the shape of Wigner-Seitz cells of body-centered cubic crystals [20].

The largest distance from the polyhedral center to its surface along (*abc*) directions, $s_{abc}(R_{100}, R_{111})$, is given by

$$s_{100}(R_{100}, R_{111}) = R_{100} \qquad (54a)$$

$$s_{110}(R_{100}, R_{111}) = \sqrt{(3/2)} \, R_{111} \qquad (54b)$$

$$s_{111}(R_{100}, R_{111}) = R_{111} \qquad (54c)$$

Further, the area of each square {100} facet is given by $F_0$ where with (52)

$$F_0 = 2 \, (\sqrt{3} \, R_{111} - R_{100})^2 \qquad (55)$$

and of each hexagonal {111} facet by $F_1$ where with (52)

$$F_1 = (3/2)\sqrt{3} \, [R_{111}{}^2 - (\sqrt{3} \, R_{111} - R_{100})^2] \qquad (56)$$

This yields the total facet surface, $F_{\mathrm{surf}}$ (sum over all facet areas) and the volume $V_{\mathrm{tot}}$ of the polyhedron according to

$$F_{\mathrm{surf}} = 6 \, F_0 + 8 \, F_1 \qquad (57)$$

$$V_{\mathrm{tot}} = ( \, 6 \, F_0 \, R_{100} + 8 \, F_1 \, R_{111} \, ) \, / \, 3 \qquad (58)$$

The surfaces of **truncated cubic** polyhedra P($\underline{R}_{\{100\}}$; $\underline{R}_{\{111\}}$) with $2 < \sqrt{3} R_{111} / R_{100} < 3$ exhibit also 6 {100} facets and 8 {111} facets as shown in Fig. 6b and there are 24 polyhedral corners which can be evaluated by methods described in Sec. S.3 of the Supplement. They are described by vectors $\underline{C}_{\{11g\}}$ relative to the center, where in Cartesian coordinates with (51)

$$\underline{C}_{\{11g\}} = R_{100} (\pm 1, \pm 1, \pm g) \ , \quad = R_{100} (\pm 1, \pm g, \pm 1) \ , \quad = R_{100} (\pm g, \pm 1, \pm 1) \tag{59}$$

$$g = \sqrt{3} \, R_{111}/R_{100} - 2 = h - 1 \ , \quad 0 \le g \le 1$$

The 6 {100} facets are of the same octagonal shape where each facet extends between eight adjacent corners $\underline{C}_{\{11g\}}$, such as $\underline{C}_{(11g)}$, $\underline{C}_{(1g1)}$, $\underline{C}_{(1-g1)}$, $\underline{C}_{(1-1g)}$, $\underline{C}_{(1-1-g)}$, $\underline{C}_{(1-g-1)}$, $\underline{C}_{(1g-1)}$, $\underline{C}_{(11-g)}$. Of the resulting eight alternating edges four connect corners, such as $\underline{C}_{(11g)}$ with $\underline{C}_{(11-g)}$, at distances $d_{f3}$ while four connect corners, such as $\underline{C}_{(11g)}$ with $\underline{C}_{(1g1)}$, at distances $d_{f4}$ given by

$$d_{f3} = (2\sqrt{3} \, R_{111} - 4R_{100}) \tag{60}$$

$$d_{f4} = \sqrt{2} \, (3R_{100} - \sqrt{3} \, R_{111}) \tag{61}$$

The 8 {111} facets are of the same equilateral triangular shape where each facet extends between three adjacent $\underline{C}_{\{11g\}}$ corners, such as $\underline{C}_{(11g)}$, $\underline{C}_{(g11)}$, $\underline{C}_{(1g1)}$. The resulting three edges connect corners, such as $\underline{C}_{(11g)}$ with $\underline{C}_{(g11)}$, at distances $d_{f4}$ according to (61).

The largest distance from the polyhedral center to its surface along (*abc*) directions, $s_{abc}(R_{100}, R_{111})$, is given by

$$s_{100}(R_{100}, R_{111}) = R_{100} \tag{62a}$$

$$s_{110}(R_{100}, R_{111}) = \sqrt{2} \, R_{100} \tag{62b}$$

$$s_{111}(R_{100}, R_{111}) = R_{111} \tag{62c}$$

Further, the area of each octagonal {100} facet is given by $F_0$ where with (51), (60), (61)

$$F_0 = 4 \, R_{100}{}^2 - 2 \, (3 \, R_{100} - \sqrt{3} \, R_{111})^2 \tag{63}$$

and of each triangular {111} facet by $F_1$ where with (61)

$$F_1 = \sqrt{3} \, (3 \, R_{100} - \sqrt{3} \, R_{111})^2/2 \tag{64}$$

This yields the total facet surface, $F_{surf}$ (sum over all facet areas) and the volume $V_{tot}$ of the polyhedron according to

$$F_{surf} = 6 \, F_0 + 8 \, F_1 \tag{65}$$

$$V_{tot} = ( \, 6 \, F_0 \, R_{100} + 8 \, F_1 \, R_{111} \, ) \, / \, 3 \tag{66}$$

There are polyhedra which can be assigned to both truncated cubic and truncated octahedral type, the *cuboctahedral* polyhedra P($\underline{R}_{\{100\}}$; $\underline{R}_{\{111\}}$) with $\sqrt{3}R_{111}/R_{100} = 2$.

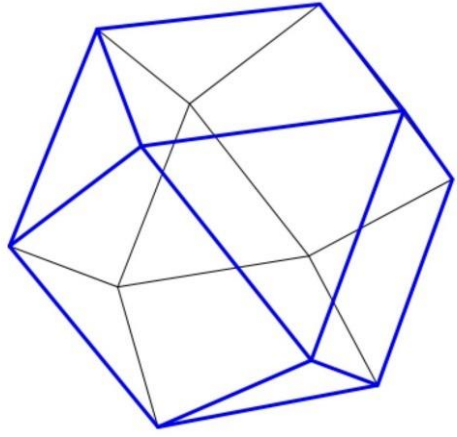

**Figure 7.** Sketch of cuboctahedral polyhedron P($\underline{R}_{\{100\}}$; $\underline{R}_{\{111\}}$) where $R_{111}/R_{100} = 1.155$ with front facets in blue and back facets in black.

These polyhedra exhibit also 6 {100} and 8 {111} facets as shown in Fig. 7 and there are 12 polyhedral corners which can be evaluated by methods described in Sec. S.3 of the Supplement. They are described by vectors $\underline{C}_{\{110\}}$ relative to the center, where in Cartesian coordinates

$$\underline{C}_{\{110\}} = R_{100} (\pm 1, \pm 1, 0), \qquad = R_{100} (\pm 1, 0, \pm 1), \qquad = R_{100} (0, \pm 1, \pm 1) \tag{67}$$

which can also be derived from (51) with $h = 1$ or from (59) with $g = 0$.

The 6 {100} facets are of the same square shape where each facet extends between four adjacent corners $\underline{C}_{\{110\}}$, such as $\underline{C}_{(110)}$, $\underline{C}_{(101)}$, $\underline{C}_{(1\text{-}10)}$, $\underline{C}_{(10\text{-}1)}$. The resulting four edges connect corners, such as $\underline{C}_{(110)}$ with $\underline{C}_{(101)}$, at distances $d_{f5}$ derived from (61) and given by

$$d_{f5} = \sqrt{2} R_{100} \tag{68}$$

The 8 {111} facets are of the same equilateral triangular shape where each facet extends between adjacent corners $\underline{C}_{\{110\}}$, such as $\underline{C}_{(110)}$, $\underline{C}_{(011)}$, $\underline{C}_{(101)}$. The resulting three edges connect corners, such as $\underline{C}_{(110)}$ with $\underline{C}_{(101)}$, at distances $d_{f5}$ according to (68).

The largest distance from the polyhedral center to its surface along ($abc$) directions, $s_{abc}(R_{100}, R_{111})$, is given by

$$s_{100}(R_{100}, R_{111}) = R_{100} \tag{69a}$$

$$s_{110}(R_{100}, R_{111}) = \sqrt{(3/2)}\, R_{111} = \sqrt{2}\, R_{100} \tag{69b}$$

$$s_{111}(R_{100}, R_{111}) = R_{111} = (2/\sqrt{3})\, R_{100} \tag{69c}$$

Further, the area of each square {100} facet is given by $F_0$ where with (68)

$$F_0 = 2\, R_{100}^2 \tag{70}$$

and of each hexagonal {111} facet by $F_1$ where with (68)

$$F_1 \; = \; (\sqrt{3})/2 \; R_{100}^2 \tag{71}$$

Thus, the total facet surface, $F_{surf}$ (sum over all facet areas) and the volume $V_{tot}$ of the polyhedron are given by

$$F_{surf} \; = \; 6 \, F_0 + 8 \, F_1 \; = \; 4 \, (3 + \sqrt{3}) \, R_{100}^2 \tag{72}$$

$$V_{tot} \; = \; (\, 6 \, F_0 \, R_{100} + 8 \, F_1 \, (2/\sqrt{3} \; R_{100}) \,) \, / \, 3 \; = \; 20/3 \; R_{100}^3 \tag{73}$$

### 3.2.3. Rhombo-octahedral Polyhedra P($\underline{R}_{\{110\}}$; $\underline{R}_{\{111\}}$)

Non-generic polyhedra P($\underline{R}_{\{110\}}$; $\underline{R}_{\{111\}}$), denoted *rhombo-octahedral*, represent intersections of two generic polyhedra, rhombohedral P($\underline{R}_{\{110\}}$) and octahedral P($\underline{R}_{\{111\}}$), see Fig. 8. If the edges of the rhombohedral polyhedron P($\underline{R}_{\{110\}}$) lie inside the octahedral polyhedron P($\underline{R}_{\{111\}}$), the resulting combination P($\underline{R}_{\{110\}}$; $\underline{R}_{\{111\}}$) will be generic rhombohedral. This requires

$$s_{111}(R_{110}) \leq s_{111}(R_{111}) \tag{74}$$

and according to (13c), (19c)

$$R_{111} \geq \sqrt{(3/2)} \; R_{110} \tag{75}$$

On the other hand, if the corners of the octahedral polyhedron P($\underline{R}_{\{111\}}$) lie inside the rhombohedral polyhedron P($\underline{R}_{\{110\}}$), the resulting combination P($\underline{R}_{\{110\}}$; $\underline{R}_{\{111\}}$) will be generic octahedral. This requires

$$s_{100}(R_{111}) \leq s_{100}(R_{110}) \tag{76}$$

and according to (13a), (19a)

$$R_{111} \leq \sqrt{(2/3)} \; R_{110} \tag{77}$$

Thus, the two generic polyhedra intersect and yield a true polyhedron P($\underline{R}_{\{110\}}$; $\underline{R}_{\{111\}}$) with both {110} and {111} facets only for facet distances $R_{110}$, $R_{111}$ with

$$\sqrt{(2/3)} \; R_{110} < R_{111} < \sqrt{(3/2)} \; R_{110} \tag{78}$$

while P($\underline{R}_{\{110\}}$; $\underline{R}_{\{111\}}$) is generic rhombohedral for $R_{111} \geq \sqrt{(3/2)} \, R_{110}$ and generic octahedral for $R_{111} \leq \sqrt{(2/3)} \, R_{110}$. As a consequence, generic polyhedra P($\underline{R}_{\{110\}}$) and P($\underline{R}_{\{111\}}$) can be described alternatively by non-generic P($\underline{R}_{\{110\}}$; $\underline{R}_{\{111\}}$) where

$$P(\underline{R}_{\{110\}}) \; = \; P(\underline{R}_{\{110\}}; \underline{R}_{\{111\}}) \; \text{with} \; R_{111} \geq \sqrt{(3/2)} \, R_{110} \quad \text{(rhombohedral)} \tag{79a}$$

$$P(\underline{R}_{\{111\}}) \; = \; P(\underline{R}_{\{110\}}; \underline{R}_{\{111\}}) \; \text{with} \; R_{110} \geq \sqrt{(3/2)} \, R_{111} \quad \text{(octahedral)} \tag{79b}$$

The surfaces of general rhombo-octahedral polyhedra P($\underline{R}_{\{110\}}$; $\underline{R}_{\{111\}}$) exhibit 12 {110} facets and 8 {111} facets as shown in Fig. 8.

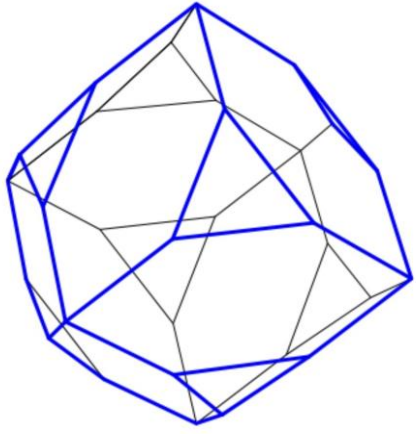

**Figure 8.** Sketch of rhombo-octahedral polyhedron P($\underline{R}_{\{110\}}$; $\underline{R}_{\{111\}}$), $R_{110}/R_{111} = 0.942$, with front facets in blue and back facets in black.

There are 30 polyhedral corners which can be evaluated by methods described in Sec. S.3 of the Supplement. They fall into two groups of 6 and 24 corners each, described by vectors $\underline{C}_{\{100\}}$ and $\underline{C}_{\{1hh\}}$ relative to the center, where in Cartesian coordinates

$$\underline{C}_{\{100\}} = \sqrt{2}\, R_{110}\, (\pm 1, 0, 0)\,, \quad = \sqrt{2}\, R_{110}\, (0, \pm 1, 0)\,, \quad = \sqrt{2}\, R_{110}\, (0, 0, \pm 1) \tag{80}$$

$$\underline{C}_{\{1hh\}} = R\, (\pm 1, \pm h, \pm h)\,, \qquad = R\, (\pm h, \pm 1, \pm h)\,, \qquad = R\, (\pm h, \pm h, \pm 1)$$

$$R = 2\sqrt{2}\, R_{110} - \sqrt{3}\, R_{111} \qquad h = (\sqrt{3}\, R_{111} - \sqrt{2}\, R_{110})\, /\, R\,, \quad 0 \le h \le 1$$

The 12 {110} facets are of the same hexagonal shape where each facet extends between six adjacent corners $\underline{C}_{\{100\}}$ and $\underline{C}_{\{1hh\}}$, such as $\underline{C}_{(100)}$, $\underline{C}_{(1hh)}$, $\underline{C}_{(hh1)}$, $\underline{C}_{(001)}$, $\underline{C}_{(h\text{-}h1)}$, $\underline{C}_{(1\text{-}hh)}$. Of the resulting six edges two connect corners, such as $\underline{C}_{(1hh)}$ with $\underline{C}_{(hh1)}$, at distances $d_{g1}$ while four connect corners, such as $\underline{C}_{(100)}$ with $\underline{C}_{(1hh)}$, at distances $d_{g2}$ given by

$$d_{g1} = \sqrt{2}\, (3\sqrt{2}\, R_{110} - 2\sqrt{3}\, R_{111}) \tag{81}$$

$$d_{g2} = \sqrt{3}\, (\sqrt{3}\, R_{111} - \sqrt{2}\, R_{110}) \tag{82}$$

The 8 triangular {111} facets are of the same equilateral triangular shape where each facet extends between adjacent corners $\underline{C}_{\{1hh\}}$, such as $\underline{C}_{(1hh)}$, $\underline{C}_{(h1h)}$, $\underline{C}_{(hh1)}$. The resulting three edges connect corners, such as $\underline{C}_{(1hh)}$ with $\underline{C}_{(hh1)}$, at distances $d_{g1}$ according to (81).

The largest distance from the polyhedral center to its surface along ($abc$) directions, $s_{abc}(R_{110}, R_{111})$, is given by

$$s_{100}(R_{110}, R_{111}) = \sqrt{2}\, R_{110} \tag{83a}$$

$$s_{110}(R_{110}, R_{111}) = R_{110} \tag{83b}$$

$$s_{111}(R_{110}, R_{111}) = R_{111} \tag{83c}$$

Further, the area of each hexagonal {110} facet is given by $F_0$ where with (80)

$$F_0 = 2\sqrt{2}\ (\sqrt{3}\ R_{111} - \sqrt{2}\ R_{110})\ (2\sqrt{2}\ R_{110} - \sqrt{3}\ R_{111}) \tag{84}$$

and of each triangular {111} facet by $F_1$ where with (81)

$$F_1 = (\sqrt{3})/2\ (3\sqrt{2}\ R_{110} - 2\sqrt{3}\ R_{111})^2 \tag{85}$$

This yields the total facet surface, $F_{surf}$ (sum over all facet areas) and the volume $V_{tot}$ of the poly-hedron according to

$$F_{surf} = 12\ F_0 + 8\ F_1 \tag{86}$$

$$V_{tot} = (\ 12\ F_0\ R_{110} + 8\ F_1\ R_{111}\ )\ /\ 3 \tag{87}$$

### 3.2.4. Cubo-rhombo-octahedral Polyhedra P($\underline{R}_{\{100\}}$; $\underline{R}_{\{110\}}$; $\underline{R}_{\{111\}}$)

Non-generic polyhedra P($\underline{R}_{\{100\}}$; $\underline{R}_{\{110\}}$; $\underline{R}_{\{111\}}$), denoted *cubo-rhombo-octahedral*, represent intersections of three generic polyhedra, cubic P($\underline{R}_{\{100\}}$), rhombohedral P($\underline{R}_{\{110\}}$), and octahedral P($\underline{R}_{\{111\}}$). Thus, they show in the most general case {100}, {110}, and {111} facets. A full dis-cussion of these polyhedra requires results for generic and non-generic polyhedra, see Secs. 3.1, and 3.2.1-3, as will be detailed in the following.

First, we consider the general notation for generic polyhedra discussed in Sec. 3.1. Cubic poly-hedra P($\underline{R}_{\{100\}}$) lie completely inside rhombohedral polyhedra P($\underline{R}_{\{110\}}$) if $R_{110} \geq \sqrt{2}\ R_{100}$ accord-ing to (31) and inside octahedral polyhedra P($\underline{R}_{\{111\}}$) if $R_{111} \geq \sqrt{3}\ R_{100}$ according to (45). This leads to polyhedra P($\underline{R}_{\{100\}}$; $\underline{R}_{\{110\}}$; $\underline{R}_{\{111\}}$) which are generic cubic where

$$\text{P}(\underline{R}_{\{100\}};\ \underline{R}_{\{110\}};\ \underline{R}_{\{111\}}) = \text{P}(\underline{R}_{\{100\}}) \quad \text{if} \quad R_{100} \leq \min(1/\sqrt{2}\ R_{110},\ 1/\sqrt{3}\ R_{111}) \tag{88}$$

The two constraints on $R_{100}$ can also be interpreted as cubic P($\underline{R}_{\{100\}}$) lying completely inside rhombo-octahedral P($\underline{R}_{\{110\}}$; $\underline{R}_{\{111\}}$).

Rhombohedral polyhedra P($\underline{R}_{\{110\}}$) lie completely inside cubic polyhedra P($\underline{R}_{\{100\}}$) if $R_{100} \geq \sqrt{2}\ R_{110}$ according to (32) and inside octahedral polyhedra P($\underline{R}_{\{111\}}$) if $R_{111} \geq \sqrt{(3/2)}\ R_{110}$ according to (75). This leads to polyhedra P($\underline{R}_{\{100\}}$; $\underline{R}_{\{110\}}$; $\underline{R}_{\{111\}}$) which are generic rhombohe-dral where

$$\text{P}(\underline{R}_{\{100\}};\ \underline{R}_{\{110\}};\ \underline{R}_{\{111\}}) = \text{P}(\underline{R}_{\{110\}}) \quad \text{if} \quad R_{110} \leq \min(1/\sqrt{2}\ R_{100},\ \sqrt{(2/3)}\ R_{111}) \tag{89}$$

The two constraints on $R_{110}$ can also be interpreted as rhombohedral P($\underline{R}_{\{110\}}$) lying completely inside cubo-octahedral P($\underline{R}_{\{100\}}$; $\underline{R}_{\{111\}}$).

Octahedral polyhedra P($\underline{R}_{\{111\}}$) lie completely inside cubic polyhedra P($\underline{R}_{\{100\}}$) if $R_{100} \geq \sqrt{3}\ R_{111}$ according to (47) and inside rhombohedral polyhedra P($\underline{R}_{\{110\}}$) if $R_{110} \geq \sqrt{(3/2)}\ R_{111}$ according to (77). This leads to polyhedra P($\underline{R}_{\{100\}}$; $\underline{R}_{\{110\}}$; $\underline{R}_{\{111\}}$) which are generic octahedral where

$$P(\underline{R}_{\{100\}}; \underline{R}_{\{110\}}; \underline{R}_{\{111\}}) \;=\; P(\underline{R}_{\{111\}}) \quad \text{if} \quad R_{111} \le \min(1/\sqrt{3}\, R_{100},\, \sqrt{(2/3)}\, R_{110}) \qquad (90)$$

The two constraints on $R_{111}$ can also be interpreted as octahedral $P(\underline{R}_{\{111\}})$ lying completely inside cubo-rhombic $P(\underline{R}_{\{100\}}; \underline{R}_{\{110\}})$.

General notations of non-generic polyhedra with two facet types, discussed in Secs. 3.2.1-3, are obtained by analogous arguments combining the previous constraints. True cubo-rhombic polyhedra $P(\underline{R}_{\{100\}}; \underline{R}_{\{110\}})$ with (34) lie completely inside octahedral polyhedra $P(\underline{R}_{\{111\}})$ if $R_{111} \ge \sqrt{3}\, R_{100}$ according to (45) and if $R_{111} \ge \sqrt{(3/2)}\, R_{110}$ according to (75). Thus,

$$P(\underline{R}_{\{100\}}; \underline{R}_{\{110\}}; \underline{R}_{\{111\}}) \;=\; P(\underline{R}_{\{100\}}; \underline{R}_{\{110\}}) \quad \text{if} \quad R_{111} \ge \min(\sqrt{3}\, R_{100},\, \sqrt{(3/2)}\, R_{110}) \qquad (91)$$

yielding with (34)

$$R_{111} \ge \sqrt{(3/2)}\, R_{110} \qquad (92)$$

True cubo-octahedral polyhedra $P(\underline{R}_{\{100\}}; \underline{R}_{\{111\}})$ with (48) lie completely inside rhombohedral polyhedra $P(\underline{R}_{\{110\}})$ if $R_{110} \ge \sqrt{2}\, R_{100}$ according to (31) and if $R_{110} \ge \sqrt{(3/2)}\, R_{111}$ according to (77). Thus,

$$P(\underline{R}_{\{100\}}; \underline{R}_{\{110\}}; \underline{R}_{\{111\}}) \;=\; P(\underline{R}_{\{100\}}; \underline{R}_{\{111\}}) \quad \text{if} \quad R_{110} \ge \min(\sqrt{2}\, R_{100},\, \sqrt{(3/2)}\, R_{111}) \qquad (93)$$

yielding with (50a), (50b), (50c)

$$R_{110} \ge \sqrt{(3/2)}\, R_{111} \quad \text{for } P(\underline{R}_{\{100\}}; \underline{R}_{\{111\}}) \text{ of truncated octahedral type} \qquad (94a)$$

$$R_{110} \ge \sqrt{2}\, R_{100} \qquad \text{for } P(\underline{R}_{\{100\}}; \underline{R}_{\{111\}}) \text{ of truncated cubic type} \qquad (94b)$$

$$R_{110} \;=\; \sqrt{2}\, R_{100} \;=\; \sqrt{(3/2)}\, R_{111} \qquad \text{for cuboctahedral } P(\underline{R}_{\{100\}}; \underline{R}_{\{111\}}) \qquad (94c)$$

True rhombo-octahedral polyhedra $P(\underline{R}_{\{110\}}; \underline{R}_{\{111\}})$ with (78) lie completely inside cubic polyhedra $P(\underline{R}_{\{100\}})$ if $R_{100} \ge \sqrt{2}\, R_{110}$ according to (32) and if $R_{100} \ge \sqrt{3}\, R_{111}$ according to (47). Thus,

$$P(\underline{R}_{\{100\}}; \underline{R}_{\{110\}}; \underline{R}_{\{111\}}) \;=\; P(\underline{R}_{\{110\}}; \underline{R}_{\{111\}}) \quad \text{if} \quad R_{100} \ge \min(\sqrt{2}\, R_{110},\, \sqrt{3}\, R_{111}) \qquad (95)$$

yielding with (78)

$$R_{100} \ge \sqrt{2}\, R_{110} \qquad (96)$$

The above constraints on $R_{abc}$ can be rephrased to treat the most general case of true cubo-rhombo-octahedral polyhedra $P(\underline{R}_{\{100\}}; \underline{R}_{\{110\}}; \underline{R}_{\{111\}})$ which exhibit {100}, {110}, as well as {111} facets. According to (34), true cubo-rhombic $P(\underline{R}_{\{100\}}; \underline{R}_{\{110\}})$ appear for

$$1/\sqrt{2}\, R_{110} < R_{100} < \sqrt{2}\, R_{110} \qquad (97)$$

Intersecting $P(\underline{R}_{\{100\}}; \underline{R}_{\{110\}})$ with $P(\underline{R}_{\{111\}})$ to yield a true polyhedron $P(\underline{R}_{\{100\}}; \underline{R}_{\{110\}}; \underline{R}_{\{111\}})$ requires according to (91) and (90)

$$\sqrt{(2/3)}\, R_{111} < R_{110} < \sqrt{(3/2)}\, R_{111} \qquad (98)$$

$$1/\sqrt{3}\, R_{100} < R_{111} < \sqrt{3}\, R_{100} \qquad (99)$$

Altogether, true polyhedra P($\underline{R}_{\{100\}}$; $\underline{R}_{\{110\}}$; $\underline{R}_{\{111\}}$) appear if constraints (97), (98), (99) are fulfilled.

Intersecting P($\underline{R}_{\{100\}}$; $\underline{R}_{\{111\}}$) with P($\underline{R}_{\{110\}}$) to yield a true polyhedron P($\underline{R}_{\{100\}}$; $\underline{R}_{\{110\}}$; $\underline{R}_{\{111\}}$) is found to yield constraints on the facet distances $R_{abc}$ which are identical with (97), (98), (99). This applies also to intersecting P($\underline{R}_{\{110\}}$; $\underline{R}_{\{111\}}$) with P($\underline{R}_{\{100\}}$). Thus, a complete analysis of the structure of true cubo-rhombo-octahedral polyhedra P($\underline{R}_{\{100\}}$; $\underline{R}_{\{110\}}$; $\underline{R}_{\{111\}}$) can be achieved by considering only one scenario where, in the following, we focus on intersecting cubo-rhombic P($\underline{R}_{\{100\}}$; $\underline{R}_{\{110\}}$) with octahedral P($\underline{R}_{\{111\}}$).

### 3.2.4.1. Polyhedra P($\underline{R}_{\{100\}}$; $\underline{R}_{\{110\}}$; $\underline{R}_{\{111\}}$) by Intersection

Intersecting cubo-rhombic P($\underline{R}_{\{100\}}$; $\underline{R}_{\{110\}}$) with octahedral P($\underline{R}_{\{111\}}$) polyhedra results in different polyhedral shapes depending on the relative sizes of the corresponding facet distances $R_{abc}$. Fixing $R_{100}$ and $R_{110}$ with (97) the shape of P($\underline{R}_{\{100\}}$; $\underline{R}_{\{110\}}$; $\underline{R}_{\{111\}}$) is fully determined by the size of its facet distance $R_{111}$. According to (97), (98), (90) $R_{111}$ must always be within the range

$$1/\sqrt{3}\ R_{100} \leq R_{111} \leq \sqrt{(3/2)}\ R_{110} \tag{100}$$

where there are two regions leading to different polyhedral shape,

outer region: $\quad R_{111}^{s} \leq R_{111} \leq \sqrt{(3/2)}\ R_{110}$ , $\quad R_{111}^{s} = 1/\sqrt{3}\ (2\sqrt{2}\ R_{110} - R_{100})$ (101)

inner region: $\quad R_{111}^{b} \leq R_{111} \leq R_{111}^{s}$ $\quad\quad R_{111}^{b} = \sqrt{(2/3)}\ R_{110}$ (102)

as illustrated in Fig. 9 by a cubo-rhombic polyhedron P($\underline{R}_{\{100\}}$; $\underline{R}_{\{110\}}$) where the dashed red triangle shows a cut along {111} indicating the boundary between the outer and inner region of corresponding cubo-rhombo-octahedral polyhedra P($\underline{R}_{\{100\}}$; $\underline{R}_{\{110\}}$; $\underline{R}_{\{111\}}$).

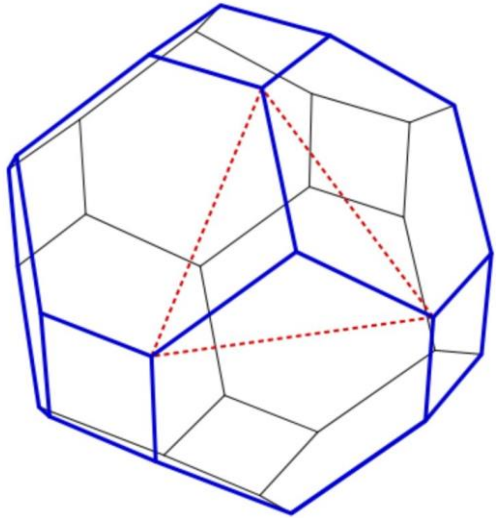

**Figure 9.** Sketch of cubo-rhombic polyhedron P($\underline{R}_{\{100\}}$; $\underline{R}_{\{110\}}$),
$R_{110}/R_{100} = 0.884$, $R_{111}/R_{100} = 1.083$, with front facets in blue and back

facets in black. The dashed red triangle shows a cut along {111} indicating the boundary between the outer and inner region of corresponding cubo-rhombo-octahedral polyhedra P($\underline{R}_{\{100\}}$; $\underline{R}_{\{110\}}$; $\underline{R}_{\{111\}}$) for $R_{111} = R_{111}^s$, $R_{111}/R_{100} = 0.866$, see text.

The **outer region** is determined by $R_{111}$ values with $R_{111}^s \leq R_{111} \leq \sqrt{(3/2)}\, R_{110}$ according to (101). Here the initial polyhedron P($\underline{R}_{\{100\}}$; $\underline{R}_{\{110\}}$) is capped at its corners $\underline{C}_{\{111\}}$ forming {111} facets. This results in 6 {100}, 12 {110}, and 8 {111} facets as shown in Fig. 10.

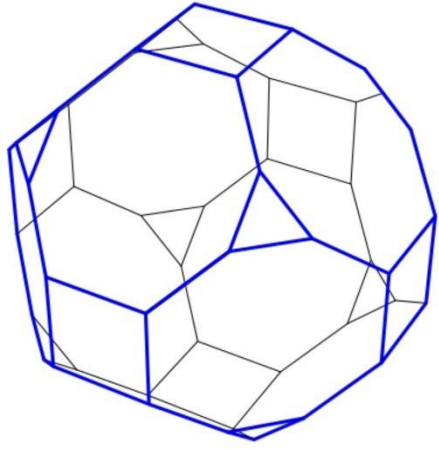

**Figure 10.** Sketch of cubo-rhombo-octahedral polyhedron P($\underline{R}_{\{100\}}$; $\underline{R}_{\{110\}}$; $\underline{R}_{\{111\}}$), $R_{110}/R_{100} = 0.884$, $R_{111}/R_{100} = 1.010$, $R_{111} \geq R_{111}^s$ (outer region), see text, with front facets in blue and back facets in black.

There are 48 polyhedral corners which can be evaluated by methods described in Sec. S.3 of the Supplement. They fall into two groups of 24 corners each, described by vectors $\underline{C}_{\{1gg\}}$ and $\underline{C}_{\{1hh\}}$ relative to the center, where in Cartesian coordinates

$$\underline{C}_{\{1gg\}} = R_{100}\,(\pm1, \pm g, \pm g)\,, \quad = R_{100}\,(\pm g, \pm1, \pm g)\,, \quad = R_{100}\,(\pm g, \pm g, \pm1)\,, \qquad (103)$$

$$\underline{C}_{\{1hh\}} = R\,(\pm1, \pm h, \pm h)\,, \quad = R\,(\pm h, \pm1, \pm h)\,, \quad = R\,(\pm h, \pm h, \pm1)$$

$$g = (\sqrt{2}\,R_{110} - R_{100})\,/\,R_{100}\,, \quad h = (\sqrt{3}\,R_{111} - \sqrt{2}\,R_{110})\,/\,R$$

$$0 \leq g \leq 1\,, \qquad 0 \leq h \leq 1\,, \qquad R = (2\sqrt{2}\,R_{110} - \sqrt{3}\,R_{111})$$

The 6 {100} facets are of the same square shape where each facet extends between four adjacent corners $\underline{C}_{\{1gg\}}$, such as $\underline{C}_{(1gg)}$, $\underline{C}_{(1-gg)}$, $\underline{C}_{(1-g-g)}$, $\underline{C}_{(1g-g)}$. The resulting four edges connect corners, such as $\underline{C}_{(1gg)}$ with $\underline{C}_{(1-gg)}$, at distances $d_{h1}$ given by

$$d_{h1} = 2\,(\sqrt{2}\,R_{110} - R_{100}) \qquad (104)$$

The 12 {110} facets are of the same octagonal shape where each facet extends between eight adjacent corners $\underline{C}_{\{1gg\}}$ and $\underline{C}_{\{1hh\}}$, such as $\underline{C}_{(1gg)}$, $\underline{C}_{(1hh)}$, $\underline{C}_{(hh1)}$, $\underline{C}_{(gg1)}$, $\underline{C}_{(g-g1)}$, $\underline{C}_{(h-h1)}$, $\underline{C}_{(1-hh)}$, $\underline{C}_{(1-gg)}$. Of the resulting eight edges two connect corners, such as $\underline{C}_{(1gg)}$ with $\underline{C}_{(1-gg)}$, at distances $d_{h1}$ given by

(104) while another two connect corners, such as $\underline{C}_{(1hh)}$ to $\underline{C}_{(hh1)}$, at distances $d_{h2}$ and four connect corners, such as $\underline{C}_{(1gg)}$ with $\underline{C}_{(1hh)}$, at distances $d_{h3}$ given by

$$d_{h2} \; = \; \sqrt{2} \; (3\sqrt{2} \; R_{110} - 2\sqrt{3} \; R_{111}) \tag{105}$$

$$d_{h3} \; = \; \sqrt{3} \; (R_{100} - 2\sqrt{2} \; R_{110} + \sqrt{3} \; R_{111}) \tag{106}$$

The 8 {111} facets are of the same equilateral triangular shape where each facet extends between adjacent corners $\underline{C}_{\{1hh\}}$, such as $\underline{C}_{(1hh)}$, $\underline{C}_{(h1h)}$, $\underline{C}_{(hh1)}$. The resulting three edges connect corners, such as $\underline{C}_{(1hh)}$ with $\underline{C}_{(hh1)}$, at distances $d_{h2}$ according to (105).

Clearly, the largest distance from the polyhedral center to its surface along (*abc*) directions, $s_{abc}(R_{100}, R_{110}, R_{111})$, equals the corresponding facet distance $R_{abc}$ i.e.

$$s_{abc}(R_{100}, R_{110}, R_{111}) \; = \; R_{abc} \tag{107}$$

Further, the area of each square {100} facet is given by $F_0$ where with (104)

$$F_0 \; = \; 4 \; (\sqrt{2} \; R_{110} - R_{100})^2 \tag{108}$$

and of each octagonal {110} facet by $F_1$ where with (103)

$$F_1 \; = \; 2\sqrt{2} \; [ \; (\sqrt{3} \; R_{111} - \sqrt{2} \; R_{110}) \; (2\sqrt{2} \; R_{110} - \sqrt{3} \; R_{111}) - (\sqrt{2} \; R_{110} - R_{100})^2 \; ] \tag{109}$$

and of each triangular {111} facet by $F_2$ where with (105)

$$F_2 \; = \; (\sqrt{3})/2 \; (3\sqrt{2} \; R_{110} - 2\sqrt{3} \; R_{111})^2 \tag{110}$$

This yields the total facet surface, $F_{surf}$ (sum over all facet areas) and the volume $V_{tot}$ of the polyhedron according to

$$F_{surf} \; = \; 6 \; F_0 + 12 \; F_1 \; + 8 \; F_2 \tag{111}$$

$$V_{tot} \; = \; ( \; 6 \; F_0 \; R_{100} + 12 \; F_1 \; R_{110} + 8 \; F_2 \; R_{111} \; ) \; / \; 3 \tag{112}$$

At the bottom of the outer region, i.e. for $R_{111} = R_{111}^s$ according to (101) the polyhedron P($\underline{R}_{\{100\}}$; $\underline{R}_{\{110\}}$; $\underline{R}_{\{111\}}$) assumes a particular shape. As before, there are 6 {100}, 12 {110}, and 8 {111} facets as shown in Fig. 11.

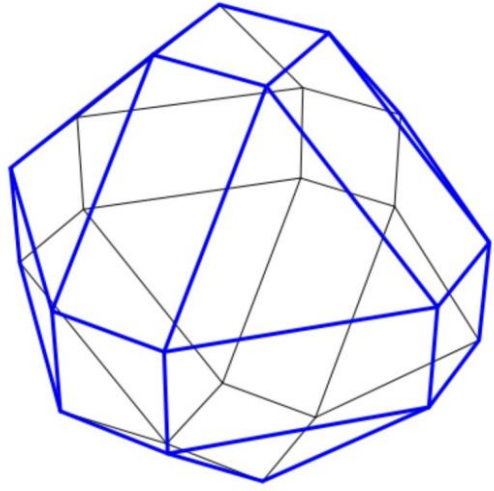

**Figure 11.** Sketch of cubo-rhombo-octahedral polyhedron
P($\underline{R}_{\{100\}}$; $\underline{R}_{\{110\}}$; $\underline{R}_{\{111\}}$), $R_{110}/R_{100} = 0.884$, $R_{111}/R_{100} = 0.866$,
$R_{111} = R_{111}^s$, see text, with front facets in blue and back facets in black.

However, the number of polyhedral corners is reduced to 24 and their Cartesian coordinates are obtained from (103) setting $R_{111} = R_{111}^s$ with (101) which yields

$$R = R_{100} \ , \qquad g = h = (\sqrt{2}\, R_{110} - R_{100}) / R_{100} \tag{113}$$

and thus

$$\underline{C}_{\{1gg\}} = R_{100} (\pm 1, \pm g, \pm g) \ , \quad = R_{100} (\pm g, \pm 1, \pm g) \ , \quad = R_{100} (\pm g, \pm g, \pm 1) \tag{114}$$

The 6 {100} facets are of the same square shape where each facet extends between four adjacent corners $\underline{C}_{\{1gg\}}$, such as $\underline{C}_{(1gg)}$, $\underline{C}_{(1-gg)}$, $\underline{C}_{(1-g-g)}$, $\underline{C}_{(1g-g)}$. The resulting four edges connect corners, such as $\underline{C}_{(1gg)}$ with $\underline{C}_{(1-gg)}$, at distances $d_{h4}$ given by

$$d_{h4} = 2 (\sqrt{2}\, R_{110} - R_{100}) \tag{115}$$

The 12 {110} facets are of the same rectangular shape where each facet extends between four adjacent corners $\underline{C}_{\{1gg\}}$, such as $\underline{C}_{(1gg)}$, $\underline{C}_{(gg1)}$, $\underline{C}_{(g-g1)}$, $\underline{C}_{(1-gg)}$. Of the resulting four edges two connect corners, such as $\underline{C}_{(1gg)}$ with $\underline{C}_{(1-gg)}$, at distances $d_{h4}$ according to (115) while two connect corners, such as $\underline{C}_{(1gg)}$ with $\underline{C}_{(gg1)}$, at distances $d_{h5}$ given by

$$d_{h5} = \sqrt{2} (2 R_{100} - \sqrt{2}\, R_{110}) \tag{116}$$

The 8 {111} facets are of the same equilateral triangular shape where each facet extends between adjacent corners $\underline{C}_{\{1gg\}}$, such as $\underline{C}_{(1gg)}$, $\underline{C}_{(g1g)}$, $\underline{C}_{(gg1)}$. The resulting three edges connect corners, such as $\underline{C}_{(1gg)}$ with $\underline{C}_{(gg1)}$, at distances $d_{h5}$ according to (116).

Clearly, the largest distance from the polyhedral center to its surface along (*abc*) directions, $s_{abc}(R_{100}, R_{110}, R_{111})$, equals the corresponding facet distance $R_{abc}$ i.e.

$$s_{abc}(R_{100}, R_{110}, R_{111}) = R_{abc} \tag{117}$$

Further, the area of each square {100} facet is given by $F_0$ where with (115)

$$F_0 = 4 (\sqrt{2}\ R_{110} - R_{100})^2 \tag{118}$$

and of each rectangular {110} facet by $F_1$ where with (115), (116)

$$F_1 = 2\sqrt{2}\ (\sqrt{2}\ R_{110} - R_{100})\ (2\ R_{100} - \sqrt{2}\ R_{110}) \tag{119}$$

and of each triangular {111} facet by $F_2$ where with (116)

$$F_2 = (\sqrt{3})/2\ (2\ R_{100} - \sqrt{2}\ R_{110})^2 \tag{120}$$

This yields the total facet surface, $F_{\text{surf}}$ (sum over all facet areas) and the volume $V_{\text{tot}}$ of the polyhedron according to

$$F_{\text{surf}} = 6\ F_0 + 12\ F_1 + 8\ F_2 \tag{121}$$

$$V_{\text{tot}} = (\ 6\ F_0\ R_{100} + 12\ F_1\ R_{110} + 8\ F_2\ R_{111}\ ) / 3 \tag{122}$$

The **inner region** is determined by $R_{111}$ values with $R_{111}^b \leq R_{111} \leq R_{111}^s$ according to (102), (101). Here the polyhedron P($\underline{R}_{\{100\}}$; $\underline{R}_{\{110\}\}}$; $\underline{R}_{\{111\}}$) with $R_{111} = R_{111}^s$ is capped further which still results in 6 {100}, 12 {110}, and 8 {111} facets, however, with changed shapes as shown in Fig. 12 and discussed below.

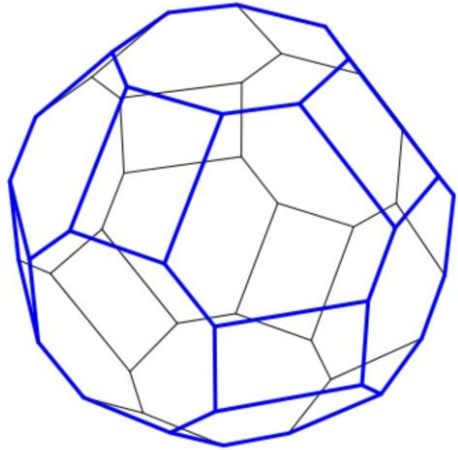

**Figure 12.** Sketch of cubo-rhombo-octahedral polyhedron
P($\underline{R}_{\{100\}}$; $\underline{R}_{\{110\}}$; $\underline{R}_{\{111\}}$), $R_{110}/R_{100} = 1.061$, $R_{111}/R_{100} = 1.010$,
$R_{111}^b \leq R_{111} \leq R_{111}^s$ (inner region), see text, with front facets in blue and
back facets in black.

There are 48 polyhedral corners which can be evaluated by methods described in Sec. S.3 of the Supplement. They are described by vectors $\underline{C}_{\{1gh\}}$ relative to the center, where in Cartesian coordinates

$$\underline{C}_{\{1gh\}} \ = \ R_{100} \ (\pm 1, \pm g, \pm h) \ , \qquad = \ R_{100} \ (\pm g, \pm 1, \pm h) \ , \qquad = \ R_{100} \ (\pm g, \pm h, \pm 1) \qquad (123)$$

$$= \ R_{100} \ (\pm 1, \pm h, \pm g) \ , \qquad = \ R_{100} \ (\pm h, \pm 1, \pm g) \ , \qquad = \ R_{100} \ (\pm h, \pm g, \pm 1)$$

$$g \ = \ (\sqrt{2} \ R_{110} - R_{100}) \ / \ R_{100} \ , \qquad h \ = \ (\sqrt{3} \ R_{111} - \sqrt{2} \ R_{110}) \ / \ R_{100}$$

$$0 \le g \le 1 \ , \qquad\qquad\qquad 0 \le h \le 1$$

The 6 {100} facets are of the same octagonal shape where each facet extends between eight adjacent corners $\underline{C}_{\{1gh\}}$, such as $\underline{C}_{(1gh)}$, $\underline{C}_{(1hg)}$, $\underline{C}_{(1-hg)}$, $\underline{C}_{(1-gh)}$, $\underline{C}_{(1-g-h)}$, $\underline{C}_{(1-h-g)}$, $\underline{C}_{(1h-g)}$, $\underline{C}_{(1g-h)}$. Of the resulting eight alternating edges four connect corners, such as $\underline{C}_{(1gh)}$ with $\underline{C}_{(1g-h)}$, at distances $d_{h6}$ while four connect corners, such as $\underline{C}_{(1gh)}$ with $\underline{C}_{(1hg)}$, at distances $d_{h7}$ given by

$$d_{h6} \ = \ 2 \ (\sqrt{3} \ R_{111} - \sqrt{2} \ R_{110}) \qquad\qquad\qquad (124)$$

$$d_{h7} \ = \ \sqrt{2} \ (2\sqrt{2} \ R_{110} - R_{100} - \sqrt{3} \ R_{111}) \qquad\qquad (125)$$

The 12 {110} facets are of the same rectangular shape where each facet extends between four adjacent corners $\underline{C}_{\{1gh\}}$, such as $\underline{C}_{(1gh)}$, $\underline{C}_{(1g-h)}$, $\underline{C}_{(g1-h)}$, $\underline{C}_{(g1h)}$. Of the resulting four alternating edges two connect corners, such as $\underline{C}_{(1gh)}$ with $\underline{C}_{(1g-h)}$, at distances $d_{h6}$ according to (124) while two connect corners, such as $\underline{C}_{(1gh)}$ with $\underline{C}_{(g1h)}$, at distances $d_{h8}$ given by

$$d_{h8} \ = \ \sqrt{2} \ (2 \ R_{100} - \sqrt{2} \ R_{110}) \qquad\qquad\qquad (126)$$

The 8 {111} facets are of the same hexagonal shape where each facet extends between six adjacent corners $\underline{C}_{\{1gh\}}$, such as $\underline{C}_{(1gh)}$, $\underline{C}_{(g1h)}$, $\underline{C}_{(h1g)}$, $\underline{C}_{(hg1)}$, $\underline{C}_{(gh1)}$, $\underline{C}_{(1hg)}$. Of the resulting six alternating edges three connect corners, such as $\underline{C}_{(1gh)}$ with $\underline{C}_{(g1h)}$, at distances $d_{h8}$ according to (126) while three connect corners, such as $\underline{C}_{(1gh)}$ with $\underline{C}_{(1hg)}$, at distances $d_{h7}$ according to (125).

Clearly, the largest distance from the polyhedral center to its surface along $(abc)$ directions, $s_{abc}(R_{100}, R_{110}, R_{111})$, equals the corresponding facet distance $R_{abc}$ i.e.

$$s_{abc}(R_{100}, R_{110}, R_{111}) \ = \ R_{abc} \qquad\qquad\qquad (127)$$

Further, the area of each square {100} facet is given by $F_0$ where with (124), (125)

$$F_0 \ = \ 4 \ (\sqrt{2} \ R_{110} - R_{100})^2 - 2 \ (2\sqrt{2} \ R_{110} - R_{100} - \sqrt{3} \ R_{111})^2 \qquad (128)$$

and of each rectangular {110} facet by $F_1$ where with (115), (116)

$$F_1 \ = \ 2\sqrt{2} \ (\sqrt{3} \ R_{111} - \sqrt{2} \ R_{110}) \ (2 \ R_{100} - \sqrt{2} \ R_{110}) \qquad\qquad (129)$$

and of each hexagonal {111} facet by $F_2$ where with (125), (126)

$$F_2 \ = \ \sqrt{3} \ [ \ (3 \ R_{110} - \sqrt{6} \ R_{111})^2 - (3/2) \ (2\sqrt{2} \ R_{110} - R_{100} - \sqrt{3} \ R_{111})^2 \ ] \qquad (130)$$

This yields the total facet surface, $F_{surf}$ (sum over all facet areas) and the volume $V_{tot}$ of the polyhedron according to

$$F_{\text{surf}} = 6\,F_0 + 12\,F_1 + 8\,F_2 \tag{131}$$

$$V_{\text{tot}} = (\,6\,F_0\,R_{100} + 12\,F_1\,R_{110} + 8\,F_2\,R_{111}\,)\,/\,3 \tag{132}$$

At the bottom of the inner region, i.e. for $R_{111} = R_{111}^b$ according to (102) the polyhedron P($\underline{R}_{\{100\}}$; $\underline{R}_{\{110\}}$; $\underline{R}_{\{111\}}$) assumes a particular shape. The 12 {110} facets of the polyhedron with $R_{111}$ in the inner region are reduced to lines and there are only 6 {100} and 8 {111} facets as shown in Fig. 13. In fact, the polyhedron is described as cubo-octahedral P($\underline{R}_{\{100\}}$; $\underline{R}_{\{111\}}$) of the truncated octahedral type ($R_{111}/R_{100} < 2/\sqrt{3}$) or of the cuboctahedral ($R_{111}/R_{100} = 2/\sqrt{3}$) type which has been discussed in detail in Sec. 3.2.2.

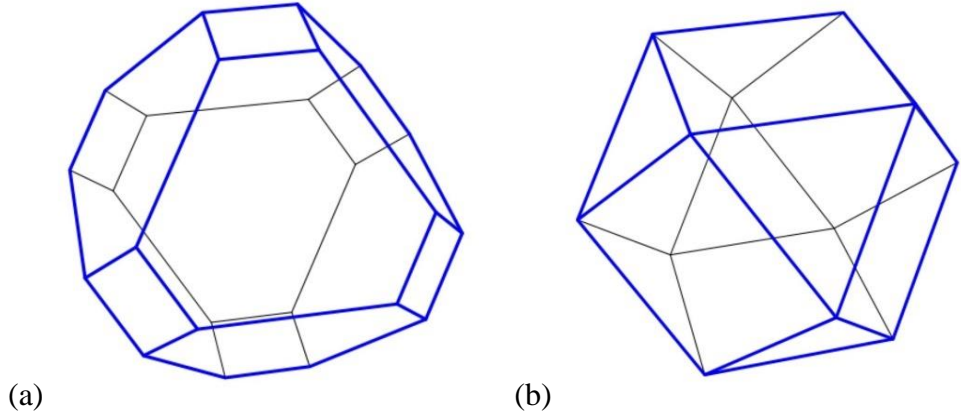

(a)                                   (b)

**Figure 13.** Sketch of cubo-rhombo-octahedral polyhedra
P($\underline{R}_{\{100\}}$; $\underline{R}_{\{110\}}$; $\underline{R}_{\{111\}}$) with $R_{111} = R_{111}^b$, see text. (a) $R_{110}/R_{100} = 0.943$,
$R_{111}/R_{100} = 0.770$, cubo-octahedral (truncated octahedral);
(b) $R_{110}/R_{100} = 1.414$, $R_{111}/R_{100} = 1.155$, cuboctahedral. Front facets are
shown in blue and back facets in black.

There are two alternative intersection procedures to achieve a true P($\underline{R}_{\{100\}}$; $\underline{R}_{\{110\}}$; $\underline{R}_{\{111\}}$) polyhedron which, however, lead to the same shapes and identical formulas for corner coordinates and all other structural parameters which have been discussed above. Therefore, they will be outlined only briefly in the following.

Intersecting cubo-octahedral P($\underline{R}_{\{100\}}$; $\underline{R}_{\{111\}}$) with rhombohedral P($\underline{R}_{\{110\}}$) polyhedra is achieved by fixing $R_{100}$ and $R_{111}$ with (99). Then the shape of P($\underline{R}_{\{100\}}$; $\underline{R}_{\{110\}}$; $\underline{R}_{\{111\}}$) is fully determined by the size of its facet distance $R_{111}$. Here we distinguish between P($\underline{R}_{\{100\}}$; $\underline{R}_{\{111\}}$) of the truncated octahedral and the truncated cubic type according to (50a), (50b). For truncated octahedral polyhedra, see Fig. 14a, and according to (97), (98), (99) $R_{110}$ must always be within the range

$$\max(\sqrt{(2/3)}\,R_{111},\,R_{100}/\sqrt{2}) \le R_{110} \le \sqrt{(3/2)}\,R_{111} \tag{133}$$

where there are two regions of different polyhedral shape,

outer region:     $R_{110}^s \leq R_{110} \leq \sqrt{(3/2)} \, R_{111}$  ,   $R_{110}^s = 1/(2\sqrt{2}) \, (\sqrt{3} \, R_{111} + R_{100})$     (134)

inner region:     $R_{110}^b \leq R_{110} \leq R_{110}^s$  ,          $R_{110}^b = \max(\sqrt{(2/3)} \, R_{111}, R_{100}/\sqrt{2})$     (135)

For truncated cubic polyhedra, see Fig. 14b, and according to (97), (98), (99) $R_{110}$ must always be within the range

$$\max(\sqrt{(2/3)} \, R_{111}, R_{100}/\sqrt{2}) \leq R_{110} \leq \sqrt{2} \, R_{111} \tag{136}$$

where there are two regions of different polyhedral shape with (134), (135),

outer region:     $R_{110}^s \leq R_{110} \leq R_{100}/\sqrt{2}$     (137)

inner region:     $R_{110}^b \leq R_{110} \leq R_{110}^s$     (138)

As an illustration, Figs. 14 show cubo-octahedral polyhedra P($\underline{R}_{\{100\}}$; $\underline{R}_{\{111\}}$) of the truncated octahedral and cubictype where the dashed red rectangles refer to cuts along {110} indicating the boundaries between the outer and inner region of corresponding cubo-rhombo-octahedral polyhedra P($\underline{R}_{\{100\}}$; $\underline{R}_{\{110\}}$; $\underline{R}_{\{111\}}$).

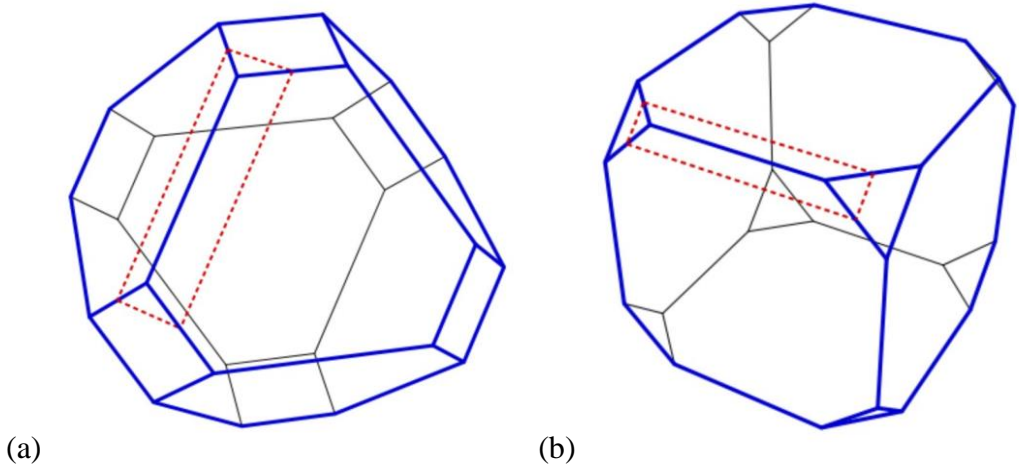

(a)                              (b)

**Figure 14.** Sketch of cubo-octahedral polyhedra P($\underline{R}_{\{100\}}$; $\underline{R}_{\{111\}}$), (a) truncated octahedral type, $R_{111}/R_{100} = 0.770$, $R_{110}/R_{100} = 0.943$, and (b) truncated cubic type, $R_{111}/R_{100} = 1.501$, $R_{110}/R_{100} = 1.414$,with front facets in blue and back facets in black. The dashed red rectangles show cuts along {110} indicating the boundaries between outer and inner regions of the corresponding cubo-rhombo-octahedral polyhedra P($\underline{R}_{\{100\}}$; $\underline{R}_{\{110\}}$; $\underline{R}_{\{111\}}$) for $R_{110} = R_{110}^s$ and (a) $R_{110}/R_{100} = 0.825$, (b) $R_{110}/R_{100} = 1.273$.

Intersecting rhombo-octahedral P($\underline{R}_{\{110\}}$; $\underline{R}_{\{111\}}$) with cubic P($\underline{R}_{\{100\}}$) polyhedra is achieved by fixing $R_{110}$ and $R_{111}$ with (98). Then the shape of P($\underline{R}_{\{100\}}$; $\underline{R}_{\{110\}}$; $\underline{R}_{\{111\}}$) is fully determined by the size of its facet distance $R_{100}$. According to (97), (98), (99) $R_{100}$ must always be within the range

$$1/\sqrt{2}\ R_{110} \leq R_{100} \leq \sqrt{2}\ R_{110} \tag{139}$$

where there are two regions leading to different polyhedral shape,

outer region: $\qquad R_{100}^s \leq R_{100} \leq \sqrt{2}\ R_{110}$ , $\qquad R_{100}^s\ =\ 2\sqrt{2}\ R_{110} - \sqrt{3}\ R_{111})$ (140)

inner region: $\qquad R_{100}^b \leq R_{100} \leq R_{111}^s$ $\qquad R_{100}^b\ =\ 1/\sqrt{2}\ R_{110}$ (141)

As an illustration, Fig. 15 shows a rhombo-octahedral polyhedron P($\underline{R}_{\{110\}}$; $\underline{R}_{\{111\}}$) where the dashed red square shows a cut along {100} indicating the boundary between the outer and inner region of corresponding cubo-rhombo-octahedral polyhedra P($\underline{R}_{\{100\}}$; $\underline{R}_{\{110\}}$; $\underline{R}_{\{111\}}$).

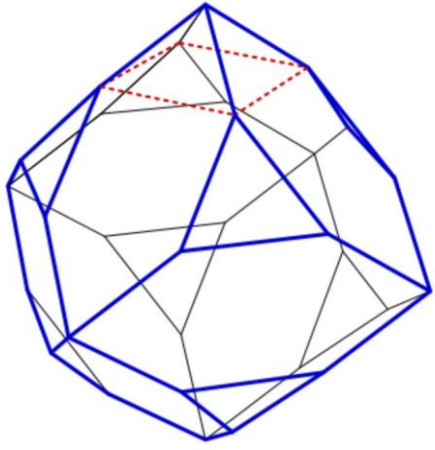

**Figure 15.** Sketch of rhombo-octahedral polyhedron P($\underline{R}_{\{110\}}$; $\underline{R}_{\{111\}}$), $R_{110}/R_{111} = 0.942$, $R_{100}/R_{111} = 1.332$, with front facets in blue and back facets in black.The dashed red square shows a cut along {100} indicating the boundary between the outer and inner region of corresponding cubo-rhombo-octahedral polyhedra P($\underline{R}_{\{100\}}$; $\underline{R}_{\{110\}}$; $\underline{R}_{\{111\}}$) for $R_{100} = R_{100}^s$, $R_{100}/R_{111} = 0.933$, see text.

### 3.2.4.2. Classification of P($\underline{R}_{\{100\}}$; $\underline{R}_{\{110\}}$; $\underline{R}_{\{111\}}$)

The discussion in Secs. 3.2.1-3 and 3.2.4.1 allows a full classification of all shapes of polyhedra P($\underline{R}_{\{100\}}$; $\underline{R}_{\{110\}}$; $\underline{R}_{\{111\}}$) with $O_h$ symmetry according to the choice of three facet distances $R_{100}$, $R_{110}$, $R_{111}$. First, we note that scaling $R_{100}$, $R_{110}$, $R_{111}$ by the same factor does not change the shape of a polyhedron. Thus, fixing $R_{100}$ at any value allows to discriminate between all shapes by considering only two parameters derived from relative facet distances $x_{110}$ and $x_{111}$ where

$$x_{110}\ =\ \sqrt{2}\ R_{110}\ /\ R_{100} \ , \qquad\qquad x_{111}\ =\ \sqrt{3}\ R_{111}\ /\ R_{100} \tag{142}$$

which leads to a two-dimensional phase diagram with $x_{110}$, $x_{111}$ as order parameters and shown in Fig. 16.

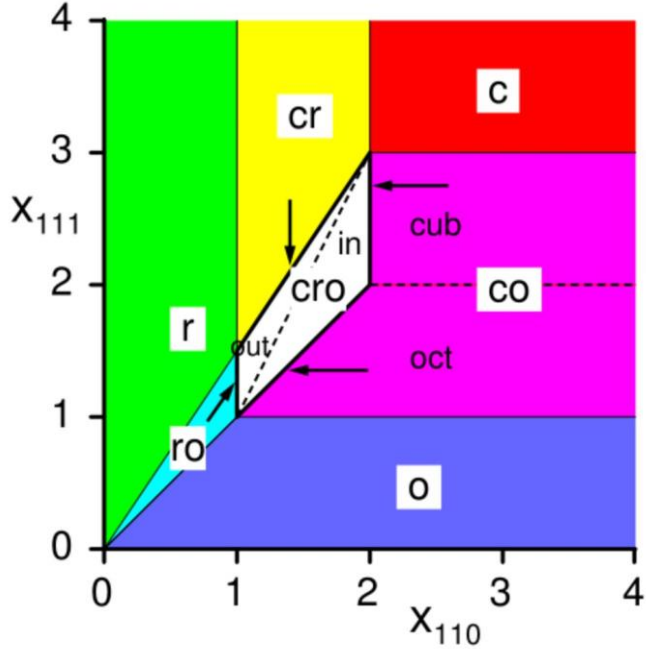

**Figure 16.** Phase diagram of all shapes of cubo-rhombo-octahedral poly-
hedra P($\underline{R}_{\{100\}}$; $\underline{R}_{\{110\}}$; $\underline{R}_{\{111\}}$) with $x_{110}$ and $x_{111}$ as order parameters. The
different phases are shown by different colors and labeled accordingly.
Arrows denote shift directions of shape identical polyhedra, see text.

True cubo-rhombo-octahedral polyhedra P($\underline{R}_{\{100\}}$; $\underline{R}_{\{110\}}$; $\underline{R}_{\{111\}}$) are defined by (97), (98), (99)
which converts to

$$1 \, < \, x_{110} \, < \, 2 \tag{143}$$

$$2/3 \, x_{111} \, < \, x_{110} \, < \, x_{111} \tag{144}$$

$$1 \, < \, x_{111} \, < \, 3 \tag{145}$$

and corresponds to the central quadrangular area labeled "cro" and confined by thick lines (cro
phase) in Fig. 16. Here the dashed line, defined according to (101) and converted to

$$x_{111} \, = \, 2 \, x_{110} \text{ - } 1 \tag{146}$$

separates polyhedra of the outer region (labeled "out") from those of the inner region (labeled
"in").

True cubo-rhombic polyhedra P($\underline{R}_{\{100\}}$; $\underline{R}_{\{110\}}$) are defined by (97) and (91) which converts to

$$1 \, < \, x_{110} \, < \, 2 \tag{147}$$

$$x_{111} \geq \min(3, \, 3/2 \, x_{110}) \, = \, 3/2 \, x_{110} \tag{148}$$

and corresponds to the infinite vertical strip labeled "cr" (cr phase) in Fig. 16. Polyhedra of this
phase along (vertical) lines of fixed $x_{110}$ differ only by the size of the octahedral polyhedron
P($\underline{R}_{\{111\}}$) outside P($\underline{R}_{\{100\}}$; $\underline{R}_{\{110\}}$). Therefore, they are identical with their counterparts at the cr/cro
phase boundary obtained by vertical shifting according to

$$(x_{110}, x_{111}) \rightarrow (x_{110}, x_{111} = 3/2 \, x_{110}) \tag{149}$$

in the phase diagram, see vertical arrow in Fig. 16.

True cubo-octahedral polyhedra P($\underline{R}_{\{100\}}$; $\underline{R}_{\{111\}}$) are defined by (99) and (93) which converts to

$$1 \, < \, x_{111} \, < \, 3 \tag{150}$$

$$x_{110} \geq \min(2 \, , x_{111}) \tag{151}$$

and corresponds to the infinite horizontal strip labeled "co" (co phase) in Fig. 16. Here the dashed line, defined according to (50c) and converted to

$$x_{111} \, = \, 2 \tag{152}$$

separates polyhedra of the truncated octahedral type ($x_{111} \leq 2$, labeled "oct") from those of the truncated cubic type ($x_{111} \geq 2$, labeled "cub"). Polyhedra of this phase along (horizontal) lines of fixed $x_{111}$ differ only by the size of the rhombohedral polyhedron P($\underline{R}_{\{110\}}$) outside P($\underline{R}_{\{100\}}$; $\underline{R}_{\{111\}}$). Therefore, they are identical with their counterparts at the co/cro phase boundary obtained by horizontal shifting according to

$$(x_{110}, x_{111}) \rightarrow (x_{110} = x_{111}, x_{111}) \qquad \text{truncated octahedral} \tag{153a}$$

$$(x_{110}, x_{111}) \rightarrow (x_{110} = 2, x_{111}) \qquad \text{truncated cubic} \tag{153b}$$

in the phase diagram, see horizontal arrows in Fig. 16.

True rhombo-octahedral polyhedra P($\underline{R}_{\{110\}}$; $\underline{R}_{\{111\}}$) are defined by (78) and (95) which converts to

$$x_{110} < x_{111} < 3/2 \, x_{110} \tag{154}$$

$$\min(x_{110}, x_{111}) \, = \, x_{110} \leq 1 \tag{155}$$

and corresponds to the triangular area labeled "ro" (ro phase) in Fig. 16. Polyhedra of this phase along (radial) lines of fixed $x_{111}/x_{110}$ differ only by the size of the cubic polyhedron P($\underline{R}_{\{100\}}$) outside P($\underline{R}_{\{110\}}$; $\underline{R}_{\{111\}}$). Therefore, they are identical with their counterparts at the ro/cro phase boundary obtained by radial shifting from the coordinate origin according to

$$(x_{110}, x_{111}) \rightarrow (x_{110} = 1, x_{111}/x_{110}) \tag{156}$$

in the phase diagram, see diagonal arrow in Fig. 16.

Generic cubic polyhedra P($\underline{R}_{\{100\}}$) are defined by (31) and (45) which converts to

$$x_{110} \geq 2 \tag{157}$$

$$x_{111} \geq 3 \tag{158}$$

and corresponds to the infinite rectangular area labeled "c" (c phase) in Fig. 16. Polyhedra of this phase are identical with their counterpart at the point

$$x_{110} = 2 , \quad x_{111} = 3 \tag{159}$$

shared between the cro and c phases since they differ only by the size of the generic rhombohedral and octahedral polyhedra outside $P(\underline{R}_{\{100\}})$.

Generic rhombohedral polyhedra $P(\underline{R}_{\{110\}})$ are defined by (32) and (75) which converts to

$$x_{100} \leq 1 \tag{160}$$

$$x_{111} \geq 3/2 \; x_{110} \tag{161}$$

and corresponds to the infinite vertical strip labeled "r" (r phase) in Fig. 16. Polyhedra of this phase are identical with their counterpart at the point

$$x_{110} = 1 , \quad x_{111} = 3/2 \tag{162}$$

shared between the cro and r phases since they differ only by the size of the generic cubic and octahedral polyhedra outside $P(\underline{R}_{\{110\}})$.

Generic octahedral polyhedra $P(\underline{R}_{\{111\}})$ are defined by (47) and (77) which converts to

$$x_{111} \leq 1 \tag{163}$$

$$x_{110} \geq x_{111} \tag{164}$$

and corresponds to the infinite horizontal strip labeled "o" (o phase) in Fig. 16. Polyhedra of this phase are identical with their counterpart at the point

$$x_{110} = 1 , \quad x_{111} = 1 \tag{165}$$

shared between the cro and o phases since they differ only by the size of the generic cubic and rhombohedral polyhedra outside $P(\underline{R}_{\{111\}})$.

Altogether, the phase diagram shown in Fig. 16 covers all possible definitions of cubo-rhombo-octahedral polyhedra $P(\underline{R}_{\{100\}}; \underline{R}_{\{110\}}; \underline{R}_{\{111\}})$ where, however, polyhedra of truly different shape are already fully accounted for by $x_{110}, x_{111}$ values inside the quadrangular area defined by (143), (144), and (145) including its edges and corners which can be described as

$$1 \leq x_{110} \leq 2 \tag{166}$$

$$x_{110} \leq x_{111} \leq 3/2 \; x_{111} \tag{167}$$

If the polyhedra are to limit nanoparticles of metal atoms, which represent sections with internal cubic bulk structure, then facet distances $R_{100}$, $R_{110}$, and $R_{111}$ assume discrete values due to the lattice periodicity. As a result, parameters $x_{110}$ and $x_{111}$ become fractional forming a homogeneous network of possible values inside the phase diagram where the network mesh is square for simple cubic (sc), rectangular 1x2 for face centered cubic (fcc), and rectangular 2x1 for body centered cubic (bcc) lattices. Further, the network mesh size decreases with increasing NP size. This is illustrated in Fig. 17 showing the phase diagram of Fig. 16 with an added network of points corresponding to an atom centered metal NP with internal face centered cubic (fcc) lattice of lattice constant $a$ and $R_{100} = 4\,a$. Here the facet distances are given by

$$R_{100} = N\,a/2\,, \quad R_{110} = M\,a/(2\sqrt{2})\,, \quad R_{111} = K\,a/\sqrt{3}\,, \qquad N, M, K \text{ integer} \tag{168}$$

and hence according to (142)

$$x_{110} = M\,/\,N\,, \qquad x_{111} = 2K\,/\,N \tag{169}$$

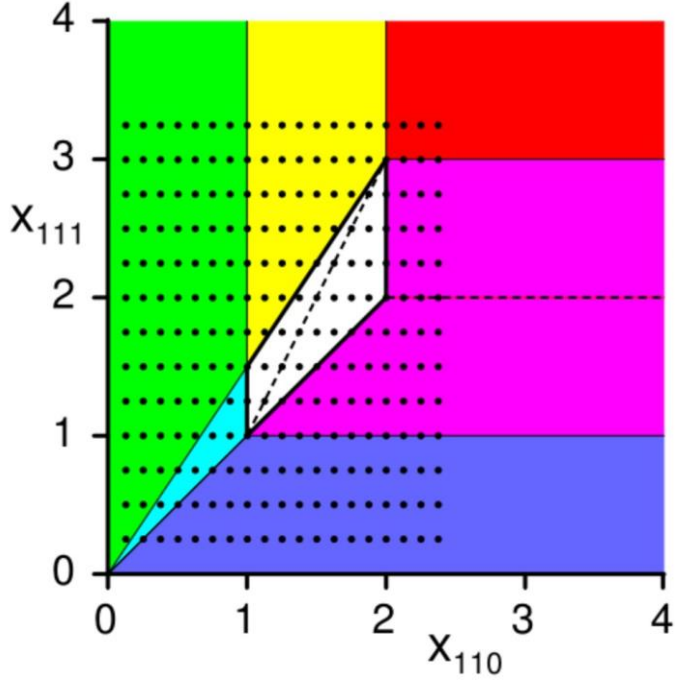

**Figure 17.** Phase diagram of all shapes of cubo-rhombo-octahedral polyhedra P($\underline{R}_{\{100\}}$; $\underline{R}_{\{110\}}$; $\underline{R}_{\{111\}}$) with $x_{110}$ and $x_{111}$ as order parameters. The different phases are shown by colors identical with those of Fig. 16. The network of points illustrates possible values of $x_{110}$, $x_{111}$ corresponding to a metal NP with internal fcc lattice and given $R_{100}$, see text.

There are two alternative classification schemes of polyhedral shapes which are mentioned only briefly and discussed in detail in Sec. S.2 of the Supplement. First, fixing $R_{110}$ at any value allows to discriminate between all shapes of polyhedra P($\underline{R}_{\{100\}}$; $\underline{R}_{\{110\}}$; $\underline{R}_{\{111\}}$) by considering two parameters derived from relative facet distances $y_{100}$ and $y_{111}$ where

$$y_{100} \,=\, R_{100}\,/\,(\sqrt{2}\,R_{110})\,, \qquad\qquad y_{111} \,=\, \sqrt{3}\,R_{111}\,/\,(\sqrt{2}\,R_{110}) \tag{170}$$

This leads to a two-dimensional phase diagram with $y_{100}$, $y_{111}$ as order parameters and shown in Fig. 18a. Second, fixing $R_{111}$ at any value allows to discriminate between all shapes by considering two parameters derived from relative facet distances $z_{100}$ and $z_{110}$ where

$$z_{100} \,=\, R_{100}\,/\,(\sqrt{3}\,R_{111})\,, \qquad\qquad z_{110} \,=\, \sqrt{2}\,R_{110}\,/\,(\sqrt{3}\,R_{111}) \tag{171}$$

This leads to a two-dimensional phase diagram with $z_{100}$, $z_{110}$ as order parameters and shown in Fig. 18b.

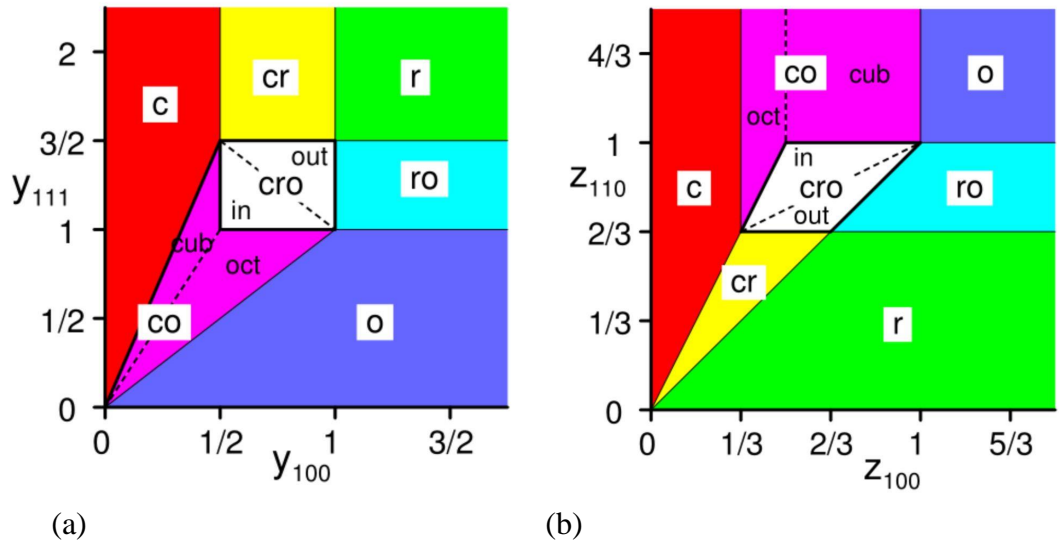

(a)                                    (b)

**Figure 18.** Phase diagram of all shapes of cubo-rhombo-octahedral polyhedra P($\underline{R}_{\{100\}}$; $\underline{R}_{\{110\}}$; $\underline{R}_{\{111\}}$); (a) with $y_{100}$ and $y_{111}$ as order parameters, (b) with $z_{100}$ and $z_{110}$ as order parameters. The different phases shown by different colors and labeled accordingly, see text.

# 4. Conclusions

The present theoretical analysis gives a full account of the shape and structure of compact polyhedra with cubic $O_h$ symmetry. The polyhedral surfaces can be described by facets representing planar sections with normal vectors along selected directions ($a$, $b$, $c$) together with their $O_h$ symmetry equivalents. Here we focus on facets reflecting normal directions of families $\{abc\}$ = $\{100\}$, = $\{110\}$, and = $\{111\}$ in Cartesian coordinates which are suggested for metal NPs with cubic lattices respresenting sections of high density monolayers of the cubic bulk. The structure evaluation identifies three types of generic polyhedra, cubic, rhombohedral, and octahedral, which can serve for the description of non-generic polyhedra as intersections of corresponding generic species. Their structural properties are shown to be fully determined by only three structure parameters, facet distances $R_{100}$, $R_{110}$, and $R_{111}$ of the three types of facets. In fact, all polyhedral shapes can already be characterized by only two relative facet distances, such as $x_{110} = R_{110}/R_{100}$ and $x_{111} = R_{111}/R_{100}$ which provides a complete phase diagram classifying the polyhedra. If the polyhedra limit nanoparticles of metal atoms, which form sections with internal cubic bulk structure, then parameters $x_{110}$ and $x_{111}$ become fractional forming a homogeneous network of possible values with square or rectangular mesh where the mesh size decreases with increasing NP size. Further, structural properties of generic polyhedra of $O_h$ symmetry, confined by facets with normal vectors of one general $\{abc\}$ family, representing up to 48 facet directions, have been analyzed. They are shown to be fully characterized by only a facet distance $R_{abc}$ and all components, $a$, $b$, $c$, of the corresponding facet normal vector family. Detailed structural properties of all polyhedra, such as shape, size, and facet surfaces, are discussed in analytical and numerical detail with visualization of characteristic examples.

Clearly, there is a multitude of other polyhedra of $O_h$ symmetry with mixtures of facets that are described by different facet directions $\{abc\}$ not accounted for in this work and having to be dealt with separately in each case. Also polyhedra of other symmetries, like hexagonal or icosahedral, can be considered. However, the present results cover already a large set of polyhedra, which allows a clear classification. The structure results may also be used as a repository assisting the interpretation of structures of real compact nanoparticles observed by experiment as well as to be used in corresponding nanoparticle simulations. Examples include mesoscopic crystallites with internal cubic lattice assuming polyhedral shape.

# S. Supplementary Information

## S.1. Special Cases of Generic Polyhedra P($\underline{R}_{\{abc\}}$)

In this Section we disuss results for generic polyhedra which derive from the generic polyhedron P($\underline{R}_{\{abc\}}$) for special values of $a$, $b$, $c$. Here we can safely assume $a \geq b \geq c \geq 0$ as shown in Sec. 3.1.4 which already accounted for polyhedra with $a > b > c > 0$, see also Fig.S0.

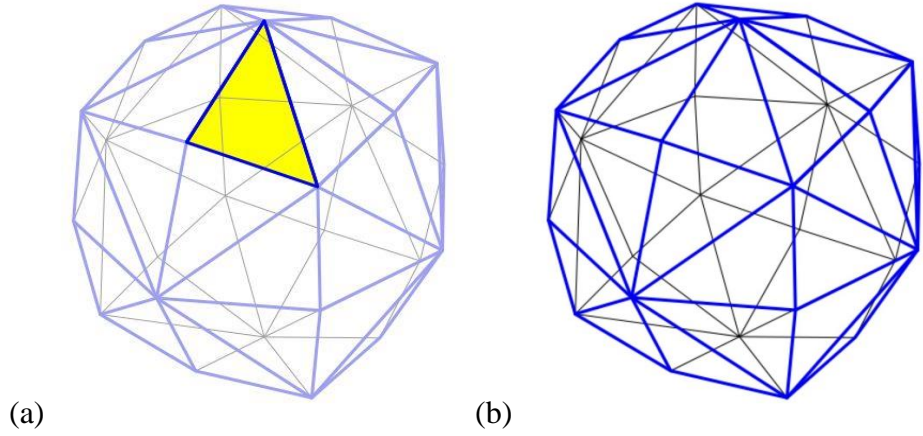

(a)                (b)

**Figure S0.** Sketch of polyhedron of type $\{abc\}$ with $a = 7$, $b = 3$, c = 1. (a) Polyhedron sketch with front facets in light blue and back facets in gray. An elementary triangular facet is emphasized in dark blue with yellow filling. (b) Full polyhedron sketch with front facets in blue and back facets in black, see text.

This leaves us with polyhedra where component values of $a$, $b$, $c$ agree with each other or are zero as discussed below.

For the following, we note that corner vectors $\underline{C}_{(100)}$, $\underline{C}_{(110)}$, and $\underline{C}_{(111)}$, defined by (25a), (25b), (25c), are connected by linear relationships where

$$\underline{C}_{(100)} = (a+b)/a \ (\underline{C}_{(110)} + \underline{C}_{(1\text{-}10)})/2 \tag{S1a}$$

$$= (a+b+c)/a \ (\underline{C}_{(111)} + \underline{C}_{(1\text{-}1\text{-}1)})/2 \tag{S1b}$$

$$\underline{C}_{(110)} = (a+b+c)/(a+b) \ (\underline{C}_{(111)} + \underline{C}_{(11\text{-}1)})/2 \tag{S1c}$$

$$= a/(a+b) \ (\underline{C}_{(100)} + \underline{C}_{(010)}) \tag{S1d}$$

$$\underline{C}_{(111)} = a/(a+b+c) \ (\underline{C}_{(100)} + \underline{C}_{(010)} + \underline{C}_{(001)}) \tag{S1e}$$

$$= (a+b)/(a+b+c) \ (\underline{C}_{(110)} + \underline{C}_{(101)} + \underline{C}_{(011)})/2 \tag{S1f}$$

### S.1.1. Polyhedra P($\underline{R}_{\{aac\}}$)

These polyhedra refer to components $a = b > c > 0$ and facet distances $R_{aac}$. Facet normal vectors $\underline{e}_{\{aac\}}$ define a family $\{aac\}$ and are obtained from the general set (2a) reducing to 24 vectors with

$$\underline{e}_{\{aac\}} = 1/w \, (\pm a, \pm a, \pm c), \; 1/w \, (\pm a, \pm c, \pm a), \; 1/w \, (\pm c, \pm a, \pm a) \tag{S2}$$
$$w = \sqrt{(2a^2 + c^2)}$$

According to (25a), (25c) possible polyhedral corners are given by

$$\underline{C}_{\{100\}} = R_{aac} \, w/a \; \underline{e}_{\{100\}} \tag{S3a}$$

$$\underline{C}_{\{111\}} = R_{aac} \, w \, \sqrt{3}/(2a+c) \; \underline{e}_{\{111\}} \tag{S3b}$$

while vector $\underline{C}_{\langle 110 \rangle}$ defined by (25b) yields with (S1d)

$$\underline{C}_{\langle 110 \rangle} = (\underline{C}_{\langle 100 \rangle} + \underline{C}_{\langle 010 \rangle})/2 \tag{S4}$$

which exemplifies that all $\underline{C}_{\langle 110 \rangle}$ are midpoints between adjacent $\underline{C}_{\{100\}}$ corners and do not qualify for corners themselves. Thus, polyhedra P($\underline{R}_{\{aac\}}$) include 14 different corners described by (S3a), (S3b).

As a result of the reduced number of corners, the 48 triangular facets discussed in Sec. 3.1.4 are described by 24 pairs of facet triangles belonging to identical facet vectors $\underline{R}_{\{aac\}}$ and joining along lines connecting corners $\underline{C}_{\{111\}}$ with $\underline{C}_{\{110\}}$. Thus, they combine to form 24 larger facets of isosceles triangle shape, see Fig. S1a, with two edges connecting corners $\underline{C}_{\{111\}}$ with $\underline{C}_{\{100\}}$ at distances $d_{s1}$ and one edge connecting two adjacent corners $\underline{C}_{\{100\}}$ at distance $d_{s2}$ where

$$d_{s1} = R_{aac} \, w \, \sqrt{[2a^2 + (a+c)^2)]} \, / \, [a \, (2a+c)] \tag{S5}$$

$$d_{s2} = 2 \, R_{aac} \, w \, / \, \sqrt{2}a \tag{S6}$$

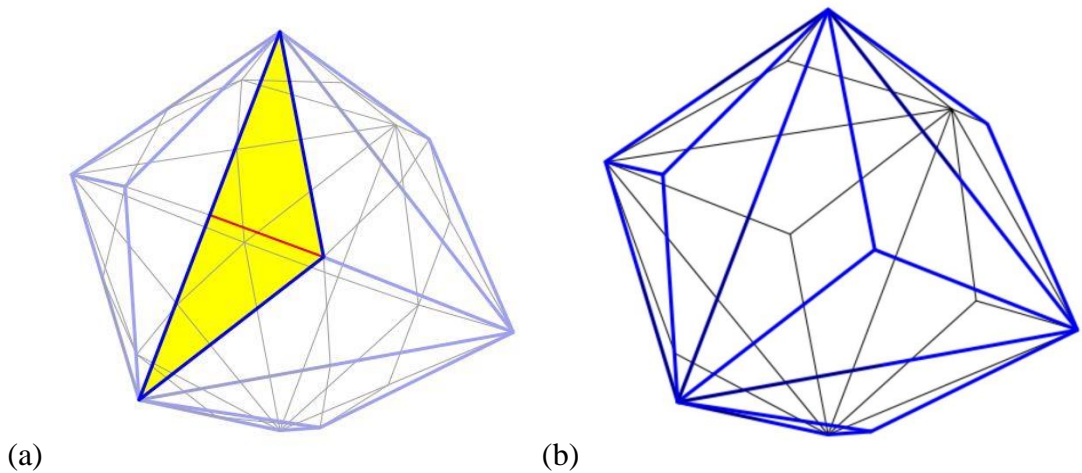

(a)                              (b)

**Figure S1.** Sketch of polyhedron of type {*aac*} with $a = b = 8$, $c = 3$. (a) Polyhedron sketch with front facets in light blue and back facets in gray, two joining coplanar triangular facets are emphasized in dark blue with yellow filling and with the separating line connecting corners $\underline{C}_{\{111\}}$ with $\underline{C}_{\{110\}}$ in red. (b) Full polyhedron sketch with front facets in blue and back facets in black, see text.

Further, the area of each facet is given by $F_0$ with

$$F_0 \;=\; |\,(\underline{C}_{(110)} - \underline{C}_{(100)}) \times (\underline{C}_{(111)} - \underline{C}_{(100)})\,| \;=\; (R_{aac}\,w)^2\,w\,/\,[2a^2\,(2a + c)] \qquad (S7)$$

Thus, the total facet surface, $F_{surf}$ (sum over all facet areas) and the volume $V_{tot}$ of the polyhedron are given by

$$F_{surf} \;=\; 24\,F_0 \;=\; 12\,(R_{aac}\,w)^2\,w\,/\,[a^2\,(2a + c)] \qquad (S8)$$

$$V_{tot} \;=\; F_{surf}\,R_{aac}\,/\,3 \;=\; 4\,(R_{aac}\,w)^3\,/\,[a^2\,(2a + c)] \qquad (S9)$$

With P($\underline{R}_{\{aac\}}$) yielding 24 facets and 14 corners the number of its polyhedral edges amounts to 36 according to (5). Fig. S1b illustrates the general polyhedron shape.

## S.1.2. Polyhedra P($\underline{R}_{\{abb\}}$)

These polyhedra refer to components $a > b = c > 0$ and facet distances $R_{abb}$. Facet normal vectors $\underline{e}_{\{abb\}}$ define a family {*abb*} and are obtained from the general set (2a) reducing to 24 vectors with

$$\underline{e}_{\{abc\}} \;=\; 1/w\,(\pm a, \pm b, \pm b),\;\; 1/w\,(\pm b, \pm a, \pm b),\;\; 1/w\,(\pm b, \pm b, \pm a) \qquad (S10)$$

$$w \;=\; \sqrt{(a^2 + 2b^2)}$$

According to (25a), (25b), (25c), possible polyhedral corners are given by

$$\underline{C}_{\{100\}} \;=\; R_{abb}\,w/a\;\underline{e}_{\{100\}} \qquad (S11a)$$

$$\underline{C}_{\{110\}} \;=\; R_{abb}\,w\,\sqrt{2}/(a+b)\;\underline{e}_{\{110\}} \qquad (S11b)$$

$$\underline{C}_{\{111\}} \;=\; R_{abb}\,w\,\sqrt{3}/(a+2b)\;\underline{e}_{\{111\}} \qquad (S11c)$$

Thus, polyhedra P($\underline{R}_{\{abb\}}$) include all 26 different corners described by (S11a), (S11b), (S11c).

As a result of the reduced number of corners, the 48 triangular facets discussed in Sec. 3.1.4 are described by 24 pairs of facet triangles belonging to identical facet vectors $\underline{R}_{\{aac\}}$ and joining along lines connecting corners $\underline{C}_{\{100\}}$ with $\underline{C}_{\{111\}}$. Thus, they combine to form 24 larger facets of quadrangle shape(kite shaped), see Fig. S2a, with two edges connecting corners $\underline{C}_{\{100\}}$ with $\underline{C}_{\{110\}}$ at distances $d_{s3}$ and two connecting $\underline{C}_{\{111\}}$ with $\underline{C}_{\{110\}}$ at distances $d_{s4}$ where

$$d_{s3} = R_{abb}\, w\, \sqrt{(a^2 + b^2)}\, /\, [a\,(a+b)] \tag{S12}$$

$$d_{s4} = R_{abb}\, w\, \sqrt{[(a+b)^2 + 2b^2)]}\, /\, [(a+b)\,(a+2b)] \tag{S13}$$

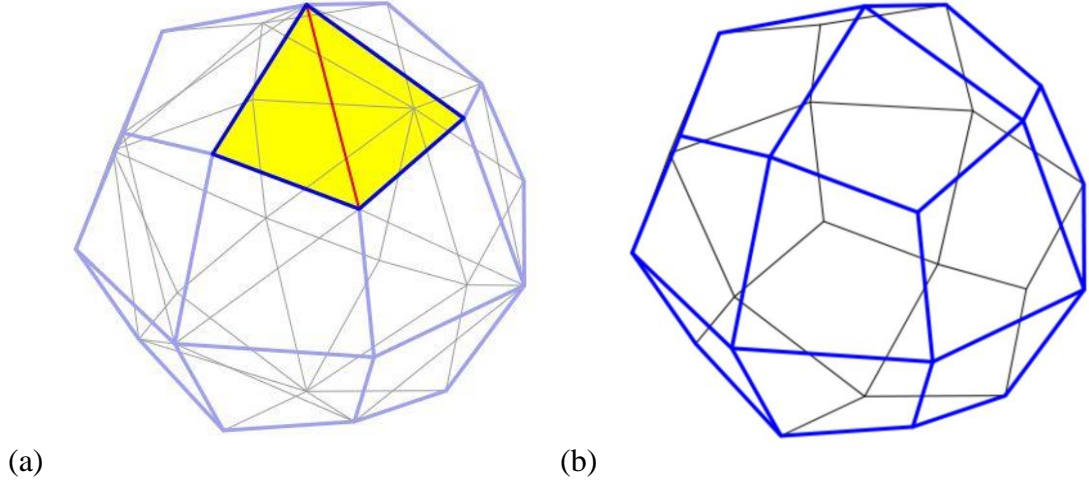

(a)                    (b)

**Figure S2.** Sketch of polyhedron of type $\{abb\}$ with $a = 9$, $b = c = 4$. (a) Polyhedron sketch with front facets in light blue and back facets in gray, two joining coplanar triangular facets are emphasized in dark blue with yellow filling and with the separating line connecting corners $\underline{C}_{\{100\}}$ with $\underline{C}_{\{111\}}$ in red. (b) Full polyhedron sketch with front facets in blue and back facets in black, see text.

Further, the area of each facet is given by $F_0$ with

$$F_0 = |\,(\underline{C}_{(110)} - \underline{C}_{(100)}) \times (\underline{C}_{(111)} - \underline{C}_{(100)})\,| = (R_{abb}\, w)^2\, w\, /\, [a\,(a+b)\,(a+2b)] \tag{S14}$$

Thus, the total facet surface, $F_{\text{surf}}$ (sum over all facet areas) and the volume $V_{\text{tot}}$ of the polyhedron are given by

$$F_{\text{surf}} = 24\, F_0 = 24\, (R_{abb}\, w)^2\, w\, /\, [a\,(a+b)\,(a+2b)] \tag{S15}$$

$$V_{\text{tot}} = F_{\text{surf}}\, R_{abb}\, /\, 3 = 8\, (R_{abb}\, w)^3\, /\, [a\,(a+b)\,(a+2b)] \tag{S16}$$

With $\mathrm{P}(\underline{R}_{\{abb\}})$ yielding 24 facets and 26 corners the number of its polyhedral edges amounts to 48 according to (5). Fig. S2b illustrates the general polyhedron shape.

### S.1.3. Polyhedra $\mathrm{P}(\underline{R}_{\{aaa\}})$

These polyhedra refer to components $a = b = c > 0$ and facet distances $R_{aaa}$. Facet normal vectors $\underline{e}_{\{aaa\}}$ define a family $\{aaa\}$ and are obtained from the general set (2a) reducing to 8 vectors with

$$\underline{e}_{\{aaa\}} = 1/w\,(\pm a, \pm a, \pm a) \qquad\qquad w = \sqrt{3}\, a \tag{S17}$$

According to (25a), (25b), (25c) possible polyhedral corners are given by

$$\underline{C}_{\{100\}} \; = \; R_{aaa} \, \sqrt{3} \; \underline{e}_{\{100\}} \tag{S18}$$

while vector $\underline{C}_{\{110\}}$ defined by (25b) yields with (S1d)

$$\underline{C}_{\{110\}} \; = \; (\underline{C}_{\{100\}} + \underline{C}_{\{010\}})/2 \tag{S19}$$

which exemplifies that all $\underline{C}_{\{110\}}$ are midpoints between adjacent $\underline{C}_{\{100\}}$ corners and do not qualify for corners themselves. Further, vector $\underline{C}_{\{111\}}$ defined by (25c) yields with (S1e) and (S1f).

$$\underline{C}_{\{111\}} \; = \; (\underline{C}_{\{100\}} + \underline{C}_{\{010\}} + \underline{C}_{\{001\}})/3 \; = \; (\underline{C}_{\{110\}} + \underline{C}_{\{101\}} + \underline{C}_{\{011\}})/3 \tag{S20}$$

which exemplifies that all $\underline{C}_{\{111\}}$ are centers of triangles between adjacent $\underline{C}_{\{100\}}$ corners as well as $\underline{C}_{\{110\}}$ and do not qualify for corners themselves. Thus, polyhedra $P(\underline{R}_{\{aaa\}})$ include all 6 different corners described by (S18).

As a result of the reduced number of corners, the 48 triangular facets discussed in Sec. 3.1.4 form 8 groups of 6 facet triangles each belonging to identical facet vectors $\underline{R}_{\{aaa\}}$ and joining along lines connecting corners $\underline{C}_{\{100\}}$ with $\underline{C}_{\{111\}}$ as well as $\underline{C}_{\{110\}}$ with $\underline{C}_{\{111\}}$. Thus, they combine to form larger facets of equilateral triangle shape, see Fig. S3a, with three edges connecting adjacent $\underline{C}_{\{100\}}$ corners at distances $d_{s5}$ where

$$d_{s5} \; = \; \sqrt{6} \, R_{aaa} \tag{S21}$$

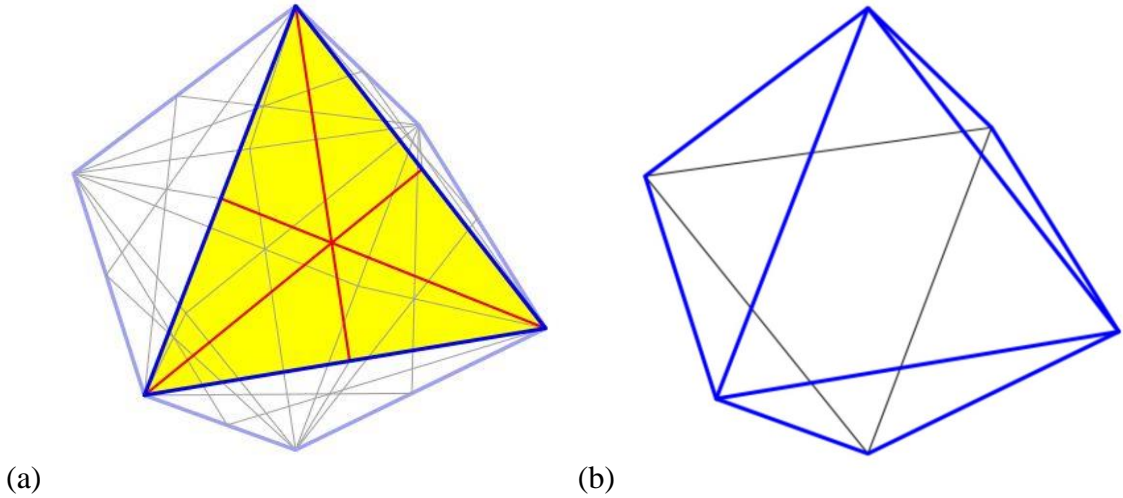

(a)                                    (b)

**Figure S3.** Sketch of a polyhedron of type $\{aaa\}$ with $a = b = c = 1$ (octahedron). (a) Polyhedron sketch with front facets in light blue and back facets in gray, six joining coplanar triangular facets are emphasized in dark blue with yellow filling and with the separating lines connecting $\underline{C}_{\{111\}}$ with $\underline{C}_{\{110\}}$ in red. (b) Full polyhedron sketch with front facets in blue and back facets in black, see text.

Further, the area of each facet is given by $F_0$ with

$$F_0 \quad = \, | \, (\underline{C}_{(110)} - \underline{C}_{(100)}) \times (\underline{C}_{(111)} - \underline{C}_{(100)}) \, | \; = \; (3/2)\sqrt{3} \; R_{aaa}^2 \qquad (S22)$$

Thus, the total facet surface, $F_{surf}$ (sum over all facet areas) and the volume $V_{tot}$ of the polyhedron are given by

$$F_{surf} \; = \; 8 \, F_0 \; = \; 12\sqrt{3} \; R_{aaa}^2 \qquad (S23)$$

$$V_{tot} \; = \; F_{surf} \, R_{aaa} \, / \, 3 \; = \; 4\sqrt{3} \; R_{aaa}^3 \qquad (S24)$$

With P($\underline{R}_{\{aaa\}}$) yielding 8 facets and 6 corners the number of its polyhedral edges amounts to 12 according to (5). In fact, the general shape of P($\underline{R}_{\{aaa\}}$), see Fig. S3b, can be characterized qualitatively as a regular octahedron discussed in Sec. 3.1.3.

### S.1.4. Polyhedra P($\underline{R}_{\{ab0\}}$)

These polyhedra refer to components $a > b > c = 0$ and facet distances $R_{ab0}$. Facet normal vectors $\underline{e}_{\{ab0\}}$ define a family $\{ab0\}$ and are obtained from the general set (2a) reducing to 24 vectors with

$$\underline{e}_{\{ab0\}} \; = \; 1/w \, (\pm a, \pm b, 0), \;\; 1/w \, (\pm b, \pm a, 0), \;\; 1/w \, (\pm a, 0, \pm b),$$
$$\qquad\qquad 1/w \, (\pm b, 0, \pm a), \;\; 1/w \, (0, \pm a, \pm b), \;\; 1/w \, (0, \pm b, \pm a) \qquad (S25)$$
$$w \; = \; \sqrt{(a^2 + b^2)}$$

According to (25a), (25c) possible polyhedral corners are given by

$$\underline{C}_{\{100\}} \; = \; R_{ab0} \, w/a \; \underline{e}_{\{100\}} \qquad (S26a)$$

$$\underline{C}_{\{111\}} \; = \; R_{ab0} \, w \, \sqrt{3}/(a+b) \; \underline{e}_{\{111\}} \qquad (S26b)$$

while vector $\underline{C}_{(110)}$ defined by (25b) yields with (S1c)

$$\underline{C}_{(110)} \; = \; (\underline{C}_{(111)} + \underline{C}_{(11\text{-}1)})/2 \qquad (S27)$$

which exemplifies that all $\underline{C}_{\{110\}}$ are midpoints between adjacent $\underline{C}_{\{111\}}$ corners and do not qualify for corners themselves. Thus, polyhedra P($\underline{R}_{\{ab0\}}$) include 14 different corners described by (S26a), (S26b).

As a result of the reduced number of corners, the 48 triangular facets discussed in Sec. 3.1.4 are described by 24 pairs of identical facet triangles joining along lines connecting corners $\underline{C}_{\{100\}}$ with $\underline{C}_{\{110\}}$ and belonging to identical facet vectors $\underline{R}_{\{ab0\}}$. Thus, they combine to form 24 larger facets of isosceles triangle shape, see Fig. S4a, with two edges connecting corners $\underline{C}_{\{100\}}$ with $\underline{C}_{\{111\}}$ at distances $d_{s6}$ and one edge connecting two adjacent corners $\underline{C}_{\{111\}}$ at distance $d_{s7}$ where

$$d_{s6} \; = \; R_{ab0} \, w \, \sqrt{(2a^2 + b^2)} \, / \, [a \, (a+b)] \qquad (S28)$$

$$d_{s7} \; = \; 2 \, R_{ab0} \, w \, / \, (a+b) \qquad (S29)$$

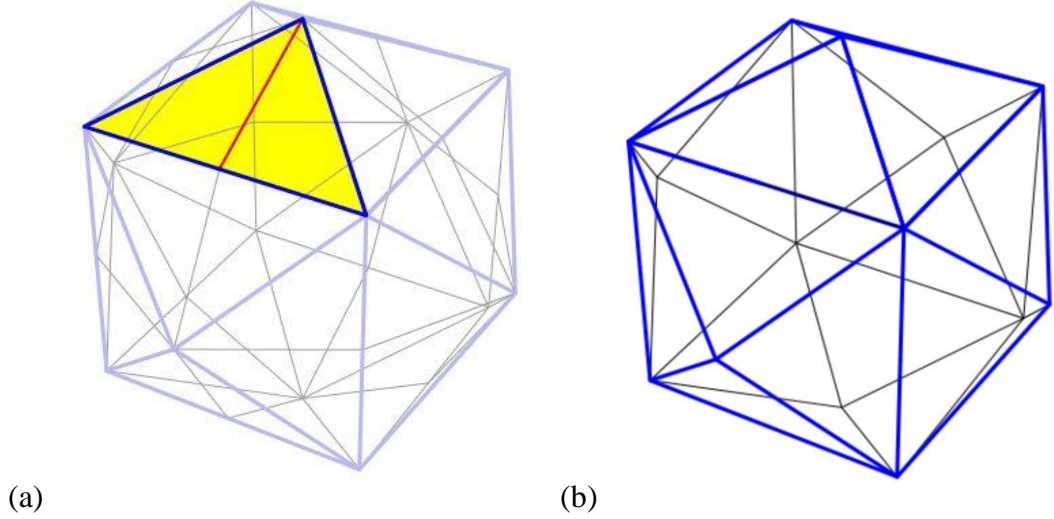

(a)                       (b)

**Figure S4.** Sketch of polyhedron of type $\{ab0\}$ with $a = 7$, $b = 4$, c = 0. (a) Polyhedron sketch with front facets in light blue and back facets in gray, two joining coplanar triangular facets are emphasized in dark blue with yellow filling and with the separating line connecting corners $\underline{C}_{\{100\}}$ with $\underline{C}_{\{110\}}$ in red. (b) Full polyhedron sketch with front facets in blue and back facets in black, see text.

Further, the area of each facet is given by $F_0$ with

$$F_0 \; = \; | \, (\underline{C}_{(110)} - \underline{C}_{(100)}) \times (\underline{C}_{(111)} - \underline{C}_{(100)}) \, | \; = \; (R_{ab0} \, w)^2 \, w \, / \, [a \, (a + b)^2] \tag{S30}$$

Thus, the total facet surface, $F_{\text{surf}}$ (sum over all facet areas) and the volume $V_{\text{tot}}$ of the polyhedron are given by

$$F_{\text{surf}} \; = \; 24 \, F_0 \; = \; 24 \, (R_{ab0} \, w)^2 \, w \, / \, [a \, (a + b)^2] \tag{S31}$$

$$V_{\text{tot}} \; = \; F_{\text{surf}} \, R_{ab0} \, / \, 3 \; = \; 8 \, (R_{ab0} \, w)^3 \, / \, [a \, (a + b)^2] \tag{S32}$$

With $\text{P}(\underline{R}_{\{ab0\}})$ yielding 24 facets and 14 corners the number of its polyhedral edges amounts to 36 according to (5). In fact, the general shape of $\text{P}(\underline{R}_{\{ab0\}})$, see Fig. S4b, can be characterized qualitatively as a cube whose six surface sides are complemented by identical square pyramids.

## S.1.5. Polyhedra $\text{P}(\underline{R}_{\{aa0\}})$

These polyhedra refer to components $a = b > c = 0$ and facet distances $R_{aa0}$. Facet normal vectors $\underline{e}_{\{aa0\}}$ define a family $\{aa0\}$ and are obtained from the general set (2a) reducing to 12 vectors with

$$\underline{e}_{\{aa0\}} \; = \; 1/w \, (\pm a, \pm a, \, 0), \;\; 1/w \, (\pm a, \, 0, \pm a), \;\; 1/w \, (0, \pm a, \pm a) \qquad w \; = \; \sqrt{2} \, a \tag{S33}$$

According to (25a), (25c) possible polyhedral corners are given by

$$\underline{C}_{\{100\}} = R_{aa0} \sqrt{2} \, \underline{e}_{\{100\}} \tag{S34a}$$

$$\underline{C}_{\{111\}} = R_{aa0} \sqrt{(3/2)} \, \underline{e}_{\{111\}} \tag{S34b}$$

while vector $\underline{C}_{(110)}$ defined by (25b) yields with (S1c)

$$\underline{C}_{(110)} = (\underline{C}_{(111)} + \underline{C}_{(11\text{-}1)})/2 \tag{S35}$$

which exemplifies that all $\underline{C}_{\{110\}}$ are midpoints between adjacent $\underline{C}_{\{111\}}$ corners and do not qualify for corners themselves. Thus, polyhedra P($\underline{R}_{\{aa0\}}$) include 14 different corners described by (S34a), (S34b).

As a result of the reduced number of corners, the 48 triangular facets discussed in Sec. 3.1.4 are described by 12 quadruplets of facet triangles belonging to identical facet vectors $\underline{R}_{\{aa0\}}$ and joining along lines connecting corners $\underline{C}_{\{100\}}$ with $\underline{C}_{\{110\}}$ as well as corners $\underline{C}_{\{111\}}$ with $\underline{C}_{\{110\}}$. Thus, they combine to form 12 larger facets of rhombic shape, see Fig. S5a, with four edges connecting corners $\underline{C}_{\{100\}}$ with $\underline{C}_{\{111\}}$ at distances $d_{s8}$ where

$$d_{s8} = \sqrt{(3/2)} \, R_{aa0} \tag{S36}$$

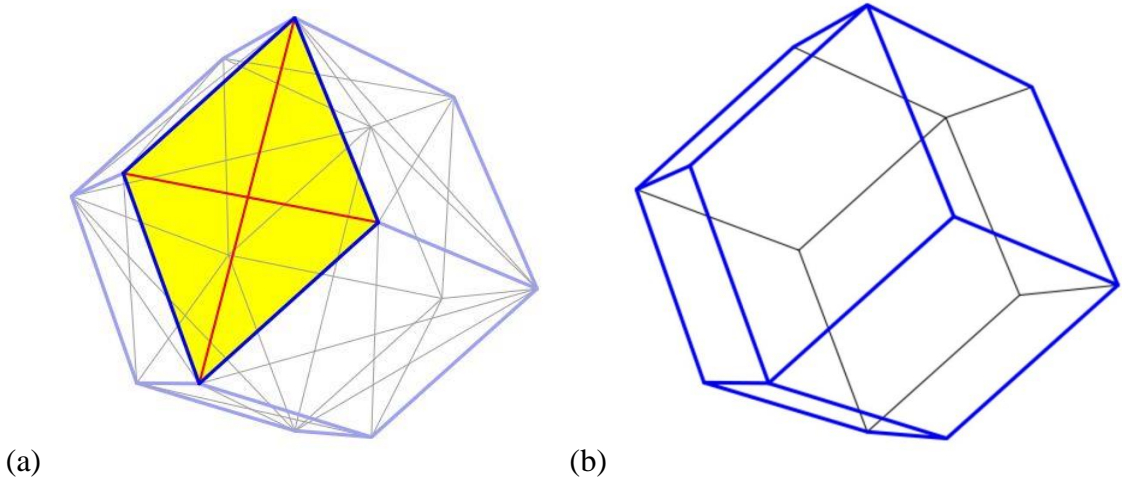

(a)                                                     (b)

**Figure S5.** Sketch of polyhedron of type $\{aa0\}$ with $a = b = 1$, c = 0 (rhombic dodecahedron). (a) Polyhedron sketch with front facets in light blue and back facets in gray, four joining coplanar triangular facets are emphasized in dark blue with yellow filling and with the separating lines connecting corners $\underline{C}_{\{100\}}$ with $\underline{C}_{\{110\}}$ and $\underline{C}_{\{111\}}$ with $\underline{C}_{\{110\}}$ in red. (b) Full polyhedron sketch with front facets in blue and back facets in black, see text.

Further, the area of each facet is given by $F_0$ with

$$F_0 = | \, (\underline{C}_{(111)} - \underline{C}_{(100)}) \times (\underline{C}_{(11\text{-}1)} - \underline{C}_{(100)}) \, | = \sqrt{2} \, R_{aa0}{}^2 \tag{S37}$$

Thus, the total facet surface, $F_{\text{surf}}$ (sum over all facet areas) and the volume $V_{\text{tot}}$ of the polyhedron are given by

$$F_{\text{surf}} = 12\, F_0 = 12\, \sqrt{2}\, R_{aa0}{}^2 \tag{S38}$$

$$V_{\text{tot}} = F_{\text{surf}}\, R_{aa0} / 3 = 4\, \sqrt{2}\, R_{aa0}{}^3 \tag{S39}$$

With P($\underline{R}_{\{aa0\}}$) yielding 12 facets and 14 corners the number of its polyhedral edges amounts to 24 according to (5). In fact, the general shape of a polyhedron P($\underline{R}_{\{aa0\}}$), see Fig. S5b, can be characterized qualitatively as a regular rhombic dodecahedron discussed already in Sec. 3.1.2.

## S.1.6. Polyhedra P($\underline{R}_{\{a00\}}$)

These polyhedra refer to components $a > b = c = 0$ and facet distances $R_{a00}$. Facet normal vectors $\underline{e}_{\{a00\}}$ define a family $\{a00\}$ and are obtained from the general set (2a) reducing to 6 vectors with

$$\underline{e}_{\{a00\}} = 1/w\, (\pm a, 0, 0),\ 1/w\, (0, \pm a, 0),\ 1/w\, (0, 0, \pm a) \qquad w = a \tag{S40}$$

According to (25c) possible polyhedral corners are given by

$$\underline{C}_{\{111\}} = R_{a00}\, \sqrt{3}\, \underline{e}_{\{111\}} \tag{S41}$$

while vector $\underline{C}_{(110)}$ defined by (25b) yields with (S1c)

$$\underline{C}_{(110)} = (\underline{C}_{(111)} + \underline{C}_{(11\text{-}1)})/2 \tag{S42}$$

which exemplifies that all $\underline{C}_{\{110\}}$ are midpoints between adjacent $\underline{C}_{\{111\}}$ corners and do not qualify for corners themselves. Further, vector $\underline{C}_{(100)}$ defined by (25a) yields with (S1a) and (25c)

$$\underline{C}_{(100)} = R_{a00}\, \underline{e}_{100} = (\underline{C}_{(111)} + \underline{C}_{(11\text{-}1)} + \underline{C}_{(1\text{-}11)} + \underline{C}_{(1\text{-}1\text{-}1)})/4 \tag{S43}$$

which exemplifies that all $\underline{C}_{\{100\}}$ are centers of squares between adjacent $\underline{C}_{\{111\}}$ corners and do not qualify for corners themselves. Thus, polyhedra P($\underline{R}_{\{a00\}}$) include 8 different corners described by (S41).

As a result of the reduced number of corners, the 48 triangular facets discussed in Sec. 3.1.4 form 6 groups of 8 facet triangles each belonging to identical facet vectors $\underline{R}_{\{a00\}}$ and joining along lines connecting $\underline{C}_{\{100\}}$ with corners $\underline{C}_{\{111\}}$. Thus, they combine to form larger facets of square shape, see Fig. S6a, with four edges connecting adjacent $\underline{C}_{\{111\}}$ corners at distances $d_{s9}$ where

$$d_{s9} = 2\, R_{a00} \tag{S44}$$

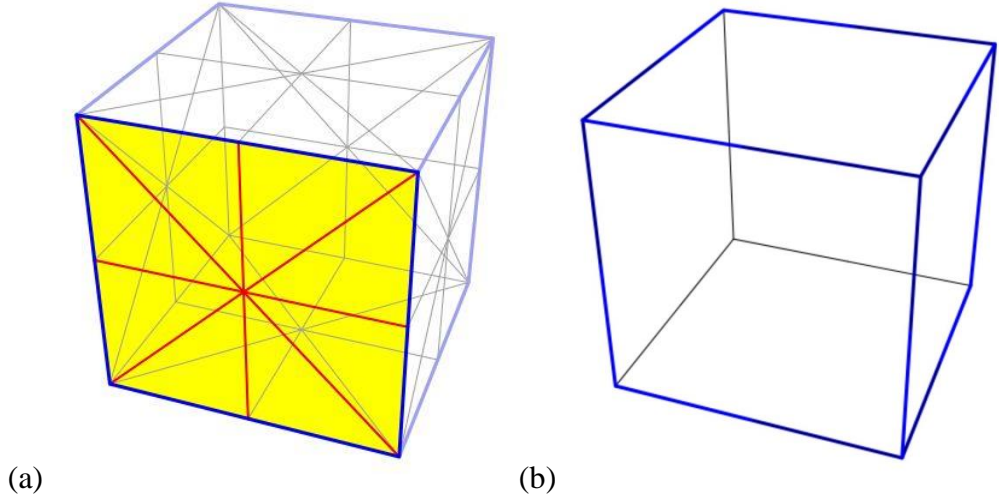

(a)                                    (b)

**Figure S6.** Sketch of polyhedron of type $\{a00\}$ with $a > b = c = 0$ (cube).
(a) Polyhedron sketch with front facets in light blue and back facets in gray, eight joining coplanar triangular facets are emphasized in dark blue with yellow filling and with separating lines connecting $\underline{C}_{\{100\}}$ with $\underline{C}_{\{110\}}$ and $\underline{C}_{\{100\}}$ with corners $\underline{C}_{\{111\}}$ in red. (b) Full polyhedron sketch with front facets in blue and back facets in black, see text.

Further, the area of each facet is given by $F_0$ with

$$F_0 \,=\, |\,(\underline{C}_{(1\text{-}11)} - \underline{C}_{(111)}) \times (\underline{C}_{(11\text{-}1)} - \underline{C}_{(111)})\,| \,=\, 4\,R_{a00}{}^2 \tag{S45}$$

Thus, the total facet surface, $F_{\mathrm{surf}}$ (sum over all facet areas) and the volume $V_{\mathrm{tot}}$ of the polyhedron are given by

$$F_{\mathrm{surf}} \,=\, 6\,F_0 \,=\, 24\,R_{a00}{}^2 \tag{S46}$$

$$V_{\mathrm{tot}} \,=\, F_{\mathrm{surf}}\,R_{a00}\,/\,3 \,=\, 8\,R_{a00}{}^3 \tag{S47}$$

With $\mathrm{P}(\underline{R}_{\{a00\}})$ yielding 6 facets and 8 corners the number of its polyhedral edges amounts to 12 according to (5). In fact, the general shape of a polyhedron $\mathrm{P}(\underline{R}_{\{a00\}})$, see Fig. S6b, can be characterized qualitatively as a cube discussed already in Sec. 3.1.1.

## S.2. Alternative classifications of $\mathrm{P}(\underline{R}_{\{100\}}; \underline{R}_{\{110\}}; \underline{R}_{\{111\}})$

There are two other classification schemes of shapes of cubo-rhombo-octahedral polyhedra $\mathrm{P}(\underline{R}_{\{100\}}; \underline{R}_{\{110\}}; \underline{R}_{\{111\}})$ which are alternatives of the scheme discussed in Sec. 3.2.4.2. This will be discussed in the following.

First, fixing $R_{110}$ at any value allows to discriminate between all polyhedral shapes by considering two parameters derived from relative facet distances $y_{100}$ and $y_{111}$ where

$$y_{100} \,=\, R_{100}\,/\,(\sqrt{2}\,R_{110}) \;,\qquad\qquad y_{111} \,=\, \sqrt{3}\,R_{111}\,/\,(\sqrt{2}\,R_{110}) \tag{S48}$$

This leads to a two-dimensional phase diagram with $y_{100}$, $y_{111}$ as order parameters and shown in Fig. S7. Note that shape identical polyhedra due to generic polyhedra which do not contribute to the polyhedral shape can be discussed analogous to Sec. 3.2.4.2.

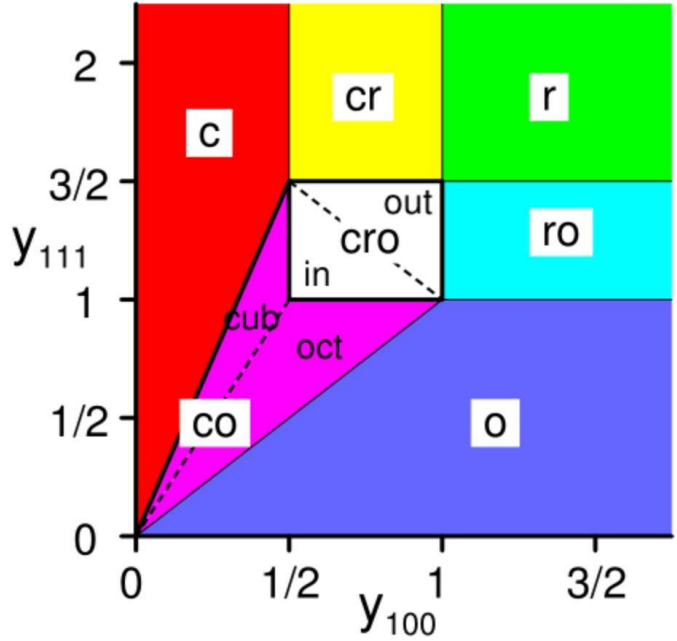

**Figure S7.** Phase diagram of all shapes of cubo-rhombo-octahedral polyhedra P($\underline{R}_{\{100\}}$; $\underline{R}_{\{110\}}$; $\underline{R}_{\{111\}}$) with $y_{100}$ and $y_{111}$ as order parameters. The different phases are shown by different colors and labeled accordingly, see text.

True cubo-rhombo-octahedral polyhedra P($\underline{R}_{\{100\}}$; $\underline{R}_{\{110\}}$; $\underline{R}_{\{111\}}$) are defined by (97), (98), (99) which converts to

$$1/2 < y_{100} < 1 \tag{S49}$$

$$1 < y_{111} < 3/2 \tag{S50}$$

$$y_{100} < y_{111} < 3\,y_{100} \tag{S51}$$

and corresponds to the central rectangular area labeled "cro" in Fig. S7. Here the dashed line, defined according to (101) and converted to

$$y_{111} = 2 - y_{100} \tag{S52}$$

separates polyhedra of the outer region (labeled "out") from those of the inner region (labeled "in").

True cubo-rhombic polyhedra P($\underline{R}_{\{100\}}$; $\underline{R}_{\{110}$) are defined by (97) and (91) which converts to

$$1/2 < y_{100} < 1 \tag{S53}$$

$$y_{111} \geq \min(3\,y_{100}, 3/2) = 3/2 \tag{S54}$$

and corresponds to the infinite vertical strip labeled "cr" in Fig. S7.

True cubo-octahedral polyhedra P($\underline{R}_{\{100\}}$; $\underline{R}_{\{111\}}$) are defined by (99) and (93) which converts to

$$y_{100} < y_{111} < 3\,y_{100} \tag{S55}$$

$$\min(2\,y_{100}, y_{111}) \leq 1 \tag{S56}$$

and corresponds to the quadrangle labeled "co" in Fig. S7. Here the dashed line, defined according to (50c) and converted to

$$y_{111} = 2\,y_{100} \tag{S57}$$

separates polyhedra of the truncated octahedral type ($y_{111} \leq 2\,y_{100}$, labeled "oct") from those of the truncated cubic type ($y_{111} \geq 2\,y_{100}$, labeled "cub").

True rhombo-octahedral polyhedra P($\underline{R}_{\{110\}}$; $\underline{R}_{\{111\}}$) are defined by (78) and (95) which converts to

$$1 < y_{111} < 3/2 \tag{S58}$$

$$y_{100} \geq \min(1, y_{111}) = 1 \tag{S59}$$

and corresponds to the infinite horizontal strip labeled "ro" in Fig. S7.

Generic cubic polyhedra P($\underline{R}_{\{100\}}$) are defined by (31) and (45) which converts to

$$y_{100} \leq 1/2 \tag{S60}$$

$$y_{111} \geq 3\,y_{100} \tag{S61}$$

and corresponds to the infinite vertical strip labeled "c" in Fig. S7.

Generic rhombohedral polyhedra P($\underline{R}_{\{110\}}$) are defined by (32) and (75) which converts to

$$y_{100} \geq 1 \tag{S62}$$

$$y_{111} \geq 3/2 \tag{S63}$$

and corresponds to the infinite rectangular area labeled "r" in Fig. S7.

Generic octahedral polyhedra P($\underline{R}_{\{111\}}$) are defined by (47) and (77) which converts to

$$y_{111} \leq y_{100} \tag{S64}$$

$$y_{111} \leq 1 \tag{S65}$$

and corresponds to the infinite horizontal strip labeled "o" in Fig. S7.

Altogether, the phase diagram shown in Fig. S7 covers all possible definitions of cubo-rhombo-octahedral polyhedra P($\underline{R}_{\{100\}}$; $\underline{R}_{\{110\}}$; $\underline{R}_{\{111\}}$) where, however, polyhedra of truly different shape are already fully accounted for by $y_{100}$, $y_{111}$ values inside the rectangular area defined by (S49), (S50), and (S51) including its edges and corners which can be described as

$$1/2 \leq y_{100} \leq 1 \tag{S66a}$$

$$1 \leq y_{111} \leq 3/2 \tag{S66b}$$

Second, fixing $R_{111}$ at any value allows to discriminate between all polyhedral shapes by considering two parameters derived from relative facet distances $z_{100}$ and $z_{110}$ where

$$z_{100} = R_{100} / (\sqrt{3} \, R_{111}) \, , \qquad\qquad z_{110} = \sqrt{2} \, R_{110} / (\sqrt{3} \, R_{111}) \tag{S67}$$

This leads to a two-dimensional phase diagram with $z_{100}$, $z_{110}$ as order parameters and shown in Fig. S8. Note that shape identical polyhedra due to generic polyhedra which do not contribute to the polyhedral shape can be discussed analogous to Sec. 3.2.4.2.

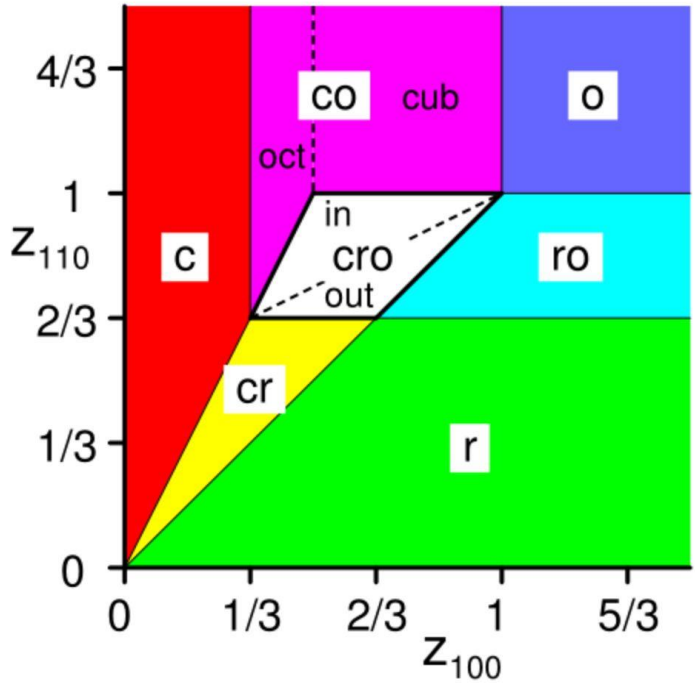

**Figure S8.** Phase diagram of all shapes of cubo-rhombo-octahedral polyhedra P($\underline{R}_{\{100\}}$; $\underline{R}_{\{110\}}$; $\underline{R}_{\{111\}}$) with $z_{100}$ and $z_{110}$ as order parameters. The different phases are shown by different colors and labeled accordingly, see text.

True cubo-rhombo-octahedral polyhedra P($\underline{R}_{\{100\}}$; $\underline{R}_{\{110\}}$; $\underline{R}_{\{111\}}$) are defined by (97), (98), (99) which converts to

$$1/2 \, z_{110} < z_{100} < z_{110} \tag{S68}$$

$$2/3 < z_{110} < 1 \qquad \text{(S69)}$$

$$1/3 < z_{100} < 1 \qquad \text{(S70)}$$

and corresponds to the central quadrangular area labeled "cro" in Fig. S8. Here the dashed line, defined according to (101) and converted to

$$z_{110} = (z_{100} + 1)/2 \qquad \text{(S71)}$$

separates polyhedra of the outer region (labeled "out") from those of the inner region (labeled "in").

True cubo-rhombic polyhedra $P(\underline{R}_{\{100\}}; \underline{R}_{110})$ are defined by (97) and (91) which converts to

$$z_{100} < z_{110} < 2\, z_{100} \qquad \text{(S72)}$$

$$\min(3\, z_{100},\, 3/2\, z_{110}) \le 1 \qquad \text{(S73)}$$

and corresponds to the triangular area labeled "cr" in Fig. S8.

True cubo-octahedral polyhedra $P(\underline{R}_{\{100\}}; \underline{R}_{\{111\}})$ are defined by (99) and (93) which converts to

$$1/3 < z_{100} < 1 \qquad \text{(S74)}$$

$$z_{110} \ge \min(2\, z_{100},\, 1) \qquad \text{(S75)}$$

and corresponds to the vertical strip labeled "co" in Fig. S8. Here the dashed line, defined according to (50c) and converted to

$$z_{100} = 1/2 \qquad \text{(S76)}$$

separates polyhedra of the truncated octahedral type ($z_{100} \le 1/2\, y_{100}$, labeled "oct") from those of the truncated cubic type ($z_{100} \ge 1/2\, y_{100}$, labeled "cub").

True rhombo-octahedral polyhedra $P(\underline{R}_{\{110\}}; \underline{R}_{\{111\}})$ are defined by (78) and (95) which converts to

$$2/3 < z_{110} < 1 \qquad \text{(S77)}$$

$$z_{100} \ge \min(z_{110},\, 1) = z_{110} \qquad \text{(S78)}$$

and corresponds to the infinite horizontal strip labeled "ro" in Fig. S8.

Generic cubic polyhedra $P(\underline{R}_{\{100\}})$ are defined by (31) and (45) which converts to

$$z_{110} \ge 2\, z_{100} \qquad \text{(S79)}$$

$$z_{100} \le 1/3 \qquad \text{(S80)}$$

and corresponds to the infinite vertical strip labeled "c" in Fig. S8.

Generic rhombohedral polyhedra $P(\underline{R}_{\{110\}})$ are defined by (32) and (75) which converts to

$$z_{110} \leq z_{100} \tag{S81}$$

$$z_{110} \leq 2/3 \tag{S82}$$

and corresponds to the infinite horizontal strip labeled "r" in Fig. S7.

Generic octahedral polyhedra P($\underline{R}_{\{111\}}$) are defined by (47) and (77) which converts to

$$z_{100} \geq 1 \tag{S83}$$

$$z_{110} \geq 1 \tag{S84}$$

and corresponds to the infinite rectangular area labeled "r" in Fig. S7.

Altogether, the phase diagram shown in Fig. S8 covers all possible definitions of cuborhombo-octahedral polyhedra P($\underline{R}_{\{100\}}$; $\underline{R}_{\{110\}}$; $\underline{R}_{\{111\}}$) where, however, polyhedra of truly different shape are already fully accounted for by $z_{100}$, $z_{110}$ values inside the rectangular area defined by (S68), (S69), and (S70) including its edges and corners which can be described as

$$z_{110} \leq z_{110} \leq 2\,z_{100} \tag{S85}$$

$$2/3 \leq z_{110} \leq 1 \tag{S86}$$

## S.3. Evaluation of Facet Edges and Corners

Vectors $\underline{R}$ pointing from the polyhedron center to a facet defined by facet vector $\underline{R}_{abc}$ are given by scalar products

$$\underline{R}_{abc}\,R \;=\; R_{abc}{}^2 \tag{S87}$$

or in Cartesian coordinates with $\underline{R} = (x, y, z)$, $\underline{R}_{abc} = (x_{abc}, y_{abc}, z_{abc})$ by

$$x_{abc}\,x + y_{abc}\,y + z_{abc}\,z \;=\; x_{abc}{}^2 + y_{abc}{}^2 + z_{abc}{}^2 \tag{S88}$$

which reflects a constraint on coordinates $x$, $y$, $z$ to yield coordinates of a two-dimensional plane.

Facet edges shared by two facets with facet vectors $\underline{R}_{abc}$, $\underline{R}_{a'b'c'}$ are described by vectors $\underline{R}_e = (x_e, y_e, z_e)$ with two constraints

$$x_{abc}\,x_e + y_{abc}\,y_e + z_{abc}\,z_e \;=\; x_{abc}{}^2 + y_{abc}{}^2 + z_{abc}{}^2 \tag{S89a}$$

$$x_{a'b'c'}\,x_e + y_{a'b'c'}\,y_e + z_{a'b'c'}\,z_e \;=\; x_{a'b'c'}{}^2 + y_{a'b'c'}{}^2 + z_{a'b'c'}{}^2 \tag{S89b}$$

to yield coordinates $x_e$, $y_e$, $z_e$ of a one-dimensional line.

Facet corners shared by three facets with facet vectors $\underline{R}_{abc}$, $\underline{R}_{a'b'c'}$, $\underline{R}_{a''b''c''}$ are described by vectors $\underline{R}_c = (x_c, y_c, z_c)$ with three constraints

$$x_{abc}\,x_c + y_{abc}\,y_c + z_{abc}\,z_c \;=\; x_{abc}{}^2 + y_{abc}{}^2 + z_{abc}{}^2 \tag{S90a}$$

$$x_{a'b'c'}\,x_c + y_{a'b'c'}\,y_c + z_{a'b'c'}\,z_c \;=\; x_{a'b'c'}{}^2 + y_{a'b'c'}{}^2 + z_{a'b'c'}{}^2 \tag{S90b}$$

$$x_{a''b''c''}\,x_c + y_{a''b''c''}\,y_c + z_{a''b''c''}\,z_c \;=\; x_{a''b''c''}{}^2 + y_{a''b''c''}{}^2 + z_{a''b''c''}{}^2 \tag{S90c}$$

representing a linear system of equations for $x_c$, $y_c$, $z_c$ defining corners. This can be written in matrix form as

$$\begin{pmatrix} x_{abc} & y_{abc} & z_{abc} \\ x_{a'b'c'} & y_{a'b'c'} & z_{a'b'c'} \\ x_{a''b''c''} & y_{a''b''c''} & z_{a''b''c''} \end{pmatrix} \cdot \begin{pmatrix} x_c \\ y_c \\ z_c \end{pmatrix} = \begin{pmatrix} R_{abc}{}^2 \\ R_{a'b'c'}{}^2 \\ R_{a''b''c''}{}^2 \end{pmatrix} \tag{S91}$$

From which corner coordinates $x$, $y$, $z$ can be evaluated according to

$$R_c = \begin{pmatrix} x_c \\ y_c \\ z_c \end{pmatrix} = \begin{pmatrix} x_{abc} & y_{abc} & z_{abc} \\ x_{a'b'c'} & y_{a'b'c'} & z_{a'b'c'} \\ x_{a''b''c''} & y_{a''b''c''} & z_{a''b''c''} \end{pmatrix}^{-1} \cdot \begin{pmatrix} R_{abc}{}^2 \\ R_{a'b'c'}{}^2 \\ R_{a''b''c''}{}^2 \end{pmatrix} \tag{S92}$$

As an example, the coordinates of a corner vector $\underline{R}_c$ shared by (100), (110), and (101) facets of P($\underline{R}_{\{100\}}$; $\underline{R}_{\{110\}}$) are given by

$$\begin{pmatrix} x_c \\ y_c \\ z_c \end{pmatrix} = \begin{pmatrix} R_{100} & 0 & 0 \\ R_{110}/\sqrt{2} & R_{110}/\sqrt{2} & 0 \\ R_{101}/\sqrt{2} & 0 & R_{101}/\sqrt{2} \end{pmatrix}^{-1} \cdot \begin{pmatrix} R_{100}{}^2 \\ R_{110}{}^2 \\ R_{101}{}^2 \end{pmatrix} = \begin{pmatrix} R_{100} \\ \sqrt{2}R_{110} - R_{100} \\ \sqrt{2}R_{101} - R_{100} \end{pmatrix} = R_{100} \begin{pmatrix} 1 \\ h \\ h \end{pmatrix} \tag{S93}$$

which exemplifies relation (36) for a corner $\underline{C}_{\{1hh\}}$ of polyhedron P($\underline{R}_{\{100\}}$; $\underline{R}_{\{110\}}$).